\documentclass[conference,compsoc,letterpaper]{IEEEtran} 
\usepackage{cite}
\usepackage{amsmath,amssymb,amsfonts}
\usepackage{makecell}  
\usepackage{tabularx}%
\usepackage{placeins}
\usepackage{algorithmic}
\usepackage{graphicx}
\usepackage{textcomp}
\usepackage{xcolor}
\usepackage{listings}
\usepackage{xcolor}
\usepackage{subcaption} 
\usepackage{enumitem} 
\usepackage{amsmath,amssymb,amsfonts}
\usepackage{algorithmic}
\usepackage{hyperref}
\usepackage{graphicx}
\usepackage{textcomp}
\usepackage{booktabs}
\usepackage{xcolor}
\usepackage{soul}
\usepackage{balance}
\usepackage{multirow}
\usepackage{dutchcal} 
\usepackage[normalem]{ulem}
\useunder{\uline}{\ul}{}
\def\BibTeX{{\rm B\kern-.05em{\sc i\kern-.025em b}\kern-.08em
    T\kern-.1667em\lower.7ex\hbox{E}\kern-.125emX}}
   
\begin{document}

\title{TempoNet: Learning Realistic Communication and Timing Patterns for Network Traffic Simulation}
\author{
\IEEEauthorblockN{
Kristen Moore\IEEEauthorrefmark{1},
Diksha Goel\IEEEauthorrefmark{1},
Cody James Christopher\IEEEauthorrefmark{1}\IEEEauthorrefmark{2},
Zhen Wang\IEEEauthorrefmark{1},\\
Minjune Kim\IEEEauthorrefmark{1},
Ahmed Ibrahim\IEEEauthorrefmark{3},
Ahmad Mohsin\IEEEauthorrefmark{3},
Seyit Camtepe\IEEEauthorrefmark{1}
}
\IEEEauthorblockA{
\IEEEauthorrefmark{1}\textit{CSIRO Data61}, Australia\\
\IEEEauthorrefmark{2}\textit{Australian National University}, Australia\\
\IEEEauthorrefmark{3}\textit{Edith Cowan University}, Australia\\
\{kristen.moore, diksha.goel, jeff.wang, minjune.kim, seyit.camtepe\}@data61.csiro.au,\\
cody.christopher@anu.edu.au,
\{ahmed.ibrahim, a.mohsin\}@ecu.edu.au
}
}

\maketitle

\begin{abstract}
Realistic network traffic simulation is critical for evaluating intrusion detection systems, stress-testing network protocols, and constructing high-fidelity environments for cybersecurity training. While attack traffic can often be layered into training environments using red-teaming or replay methods, generating authentic benign background traffic remains a core challenge—particularly in simulating the complex temporal and communication dynamics of real-world networks. This paper introduces \textbf{TempoNet}, a novel generative model that combines multi-task learning with multi-mark temporal point processes to jointly model inter-arrival times and all packet- and flow-header fields. TempoNet captures fine-grained timing patterns and higher-order correlations such as host-pair behavior and seasonal trends, addressing key limitations of GAN-, LLM-, and Bayesian-based methods that fail to reproduce structured temporal variation. TempoNet produces temporally consistent, high-fidelity traces, validated on real-world datasets. Furthermore, we show that intrusion detection models trained on TempoNet-generated background traffic perform comparably to those trained on real data, validating its utility for real-world security applications.
\end{abstract}


\maketitle

\section{Introduction}

Synthetic network traffic generation is essential for evaluating and optimizing network-based systems in applications such as performance benchmarking, intrusion detection, and cybersecurity training. However, access to real-world network traces is often limited due to privacy, proprietary, and operational security constraints, making it difficult to use authentic data in research or simulation environments. Synthetic data generation offers a promising alternative—allowing researchers to emulate realistic behavior without exposing sensitive information.

Yet without accurate background traffic simulation, Red vs. Blue exercises, cyber deception platforms, and intrusion detection evaluations risk relying on unrealistic assumptions. In cyber range training, for example, the absence of legitimate user activity means blue teams can trivially detect red team attacks by simply observing any traffic at all. As highlighted by Mäses et al.~\cite{mases2021success}, the lack of authentic background traffic leads to flawed assessments and diminished training outcomes. Similarly, deception platforms require diverse, natural-looking background flows to sustain adversary engagement and prevent detection of honeypots  (e.g., fake user login sessions, or periodic DNS queries). Replaying NetFlow or PCAP traces is brittle, inflexible, and often avoided due to privacy risks and residual metadata leakage~\cite{paxson1997end}—even anonymized traffic can reveal sensitive host or behavioral information through timing patterns and communication structures~\cite{coull2007playing}.

A key challenge lies in generating traffic that accurately replicates the complex temporal and communication dynamics of real-world environments. Recent ML-based approaches can produce high-fidelity header data but often struggle to model the multidimensional temporal dependencies present in packet-header (e.g., PCAP) and flow-header (e.g., NetFlow) traces. Generative Adversarial Networks (GANs), for example, are known to struggle with long-range dependencies, and modeling inter-arrival times is explicitly considered out of scope in state-of-the-art GAN-based methods~\cite{yin2022practical, lin2020using}. While diffusion and VAE-based models have shown promise in other domains, they remain underexplored for full-field network traffic simulation due to high inference costs and limited support for multi-field tabular data. Pre-trained Large Language Models (LLMs) have recently been adapted for structured data generation, but remain largely untested in capturing higher-order temporal correlations~\cite{borisov2022language}. Temporal Point Processes (TPPs), by contrast, offer a natural framework for modeling time-dependent event sequences, yet existing TPP-based generators are typically limited to producing only one or two event attributes~\cite{6782285, moore2016analysis, saha2019learning, price2020nonparametric}, falling short of the multidimensional realism required for comprehensive traffic simulation.

As a result, synthetic traces often lack the variability and nuance needed for security applications: blue teams and IDS tools trained in these environments can more easily distinguish simulated from real events, undermining both detection effectiveness and training value~\cite{lopes2023network, de2023unsupervised, 8171733}. Capturing these complex patterns is essential for robust benchmarking, intrusion detection, cyber range training, and deception systems that aim to reflect operational realities.

This paper introduces \textbf{TempoNet}, a novel generative framework that combines temporal point processes (TPPs) with multi-task learning to generate realistic, temporally consistent packet- and flow-header traces. TempoNet jointly models all header fields—including categorical values (e.g., IP addresses, ports), continuous variables (e.g., inter-arrival times, packet sizes), and discrete counts (e.g., flow duration)—while preserving the interdependencies across them. At its core, the model uses a recurrent neural network (RNN) to encode recent history and a log-normal mixture model TPP to sample inter-arrival times. Task-specific output layers then decode each header field, enabling multi-mark generation within a unified architecture.

TempoNet enables efficient, high-fidelity simulation of network traffic that closely mirrors real-world behavior, capturing both fine-grained timing and higher-order communication patterns such as host-pair burstiness and structured periodicity. These capabilities make it particularly suited for security-critical applications, including adversarial model evaluation, cyber deception, and blue team training.

Our main contributions include:\\
\noindent 1. \textbf{Integrated Generative Model}: A novel generative architecture that combines multi-task learning with multi-mark temporal point processes to jointly model inter-arrival times and packet- and flow-header fields, capturing structured temporal patterns such as daily and weekly seasonality.\\
\noindent 2. \textbf{Comprehensive Empirical Evaluation}: The first rigorous comparison of GAN-, LLM-, Bayesian-, and TPP-based approaches for modeling temporal and higher-order network communication dynamics. TempoNet outperforms prior methods across fidelity, diversity, and compliance metrics, and handles heterogeneous header fields including numeric, categorical, and temporally structured features such as inter-arrival times with daily and weekly seasonality.\\
\noindent 3. \textbf{Real-World Security Application}: Demonstrate the practical utility of TempoNet-generated traffic through an intrusion detection task, using real CIDDS attack data layered on top of synthetic benign traffic.

\section{Background: Temporal Point Processes}\label{ifl-lognormmix}
Temporal point processes (TPP) are probabilistic models designed to handle event data that include timestamps. Essentially, a TPP is a stochastic process where the output consists of the event times $\{t_i\}$, or equivalently, the inter-arrival times $\{ \tau_i:=t_i - t_{i-1}\}$ between events. These processes offer an intuitive framework for modeling streams of discrete events over time, employing probabilistic distributions.
The most basic TPP is the Poisson Process \cite{palm1943inten}, which assumes that events occur independently and that the time between events follows an exponential distribution.  Numerous extensions of the Poisson Process have been developed to capture more complex dependencies between event occurrences, including the Hawkes Process~\cite{10.2307/2334319}, a self-exciting process where past events temporarily increase the likelihood of future events.

A TPP is uniquely characterized by its \textit{conditional intensity function}, which specifies the rate at which events occur based on the event history. Conditional intensity functions offer a straightforward method to define TPPs with predetermined behavior. However, choosing an appropriate intensity function presents a significant challenge. Simple functions with closed-form log-likelihoods often lack expressiveness, while more complex functions require approximation methods that can lead to noisy gradient estimates during training.
An alternative approach, proposed by Shchur et al.~\cite{shchur2019intensity}, directly models the conditional probability density function of inter-event times. This intensity-free approach uses normalizing flows to create flexible TPP models that can approximate any probability density arbitrarily well, including multi-modal ones. Specifically, their LogNormMix model employs a log-normal mixture distribution, offering both flexibility and closed-form sampling capabilities.

A marked TPP builds on the standard TPP by attaching extra information, called marks, to each event. These marks can include details like the event's size, type, or other characteristics. Unlike standard TPPs, a marked TPP predicts both when events happen and the marks associated with them, allowing for a more detailed understanding of event patterns.
However, adding marks significantly increases model complexity and adds to computational demands. Careful design is required to capture how marks relate to each other and to the timing of events. Because of these challenges, existing work in the TPP domain has focused on the simpler task of modeling only one or two event attributes (marks) alongside the event timestamps.

\section{Preliminaries}
\subsection{Problem formulation}
Our input data comprises network traces in two forms: packet headers or flow headers. Packet header traces include details such as source and destination IP addresses, ports and protocol, along with associated metrics like timestamps and packet sizes. For flow header traces, each entry contains the IP 5-tuple (source and destination IPs, ports, and protocol) along with related measurements, including flow start and end times and total bytes transferred.

Our objective is to design a generative model for PCAP and NetFlow traces that achieves high fidelity to the original data, as defined by expert-informed criteria and specific application needs. TempoNet, like most prior work (e.g.,~\cite{xu2021stan, yin2022practical, lin2019generating, schoen2024tale}), models only these header-level features and does not attempt to generate payload contents. This design choice is intentional and aligns with the structure of common benchmarks and training environments, such as LANL, OpTC, and GHOSTS, which provide header-only data due to privacy and compliance considerations. Payload modeling introduces additional complexity and remains far less common in practice, as it raises complex privacy, application-specific, and modeling challenges.

Instead, TempoNet prioritizes temporal and structural fidelity—dimensions that are underexplored but critically important for evaluating detection models and simulating realistic background traffic. This focus allows our method to remain broadly applicable across diverse environments where payload access is restricted or unavailable.

The evaluation framework includes four key criteria, inspired by recent work~\cite{schoen2024tale}:

\textbf{Realism:} Measures how well the synthetic traffic matches the statistical characteristics of real data. We use Earth Mover’s Distance (EMD) for numerical fields (e.g., inter-arrival times) and Jensen-Shannon Divergence (JSD) for categorical fields (e.g., IP addresses, ports). To assess inter-feature relationships, we use Pairwise Correlation Difference (PCD), defined as the $L^2$ norm between correlation matrices of numerical features, and Contingency Matrix Difference (CMD), which compares categorical feature dependencies.

\textbf{Diversity:} Capturing the variability in the data by ensuring the generated traffic maintains a similar level of variance to the real data, assessed using Coverage and Density metrics.

\textbf{Novelty:} Verifying that synthetic samples are not direct copies of the training data, evaluated using metrics like Membership Disclosure.

\textbf{Compliance:} Ensuring adherence to network protocol rules through a Domain Knowledge Check (DKC), which tests the validity of generated traffic against known protocol constraints.

Through this multifaceted evaluation, we aim to develop a model that produces realistic, diverse, and compliant network flow traffic while effectively capturing the intrinsic correlations between features.

\subsection{Why TPPs?} Although many machine learning-based approaches treat network traffic generation as a time series problem, the underlying process governing packet and flow inter-arrival times is more accurately represented by Temporal Point Processes (TPPs). Recognizing this, several studies have adopted TPPs to demonstrate superior performance in generating, forecasting, and detecting anomalies in traffic timing and volume~\cite{6782285, moore2016analysis, saha2019learning, price2020nonparametric}.
The accurate representation of temporal dynamics in simulated background traffic is crucial for creating realistic cyber range exercises. It creates an authentic environment that mirrors real-world network behavior, increasing the complexity and educational value of exercises. Moreover, they enable effective testing of network intrusion detection and anomaly detection systems, which often rely heavily on temporal cues. 

Despite their importance, temporal dynamics are often considered out of scope by state-of-the-art network packet trace generation approaches~\cite{yin2022practical, schoen2024tale, kholgh2023pac}. Many existing methods focus on replicating distributions of static attributes, such as IPs, ports, and packet sizes, while neglecting the complex temporal dependencies present in real network traffic. This limitation underscores the need for TPP-based approaches, which are inherently designed to model temporal dynamics alongside other traffic attributes. 

However, whilst existing TPP approaches excel at capturing temporal patterns, they often overlook the modeling of other critical header trace fields, such as source and destination IP addresses, ports, protocol, and packet sizes, leaving room for improvement in generating fully representative network traffic.

When designing TPP models, three key attributes are crucial~\cite{ijcai2021p623}:

i) Flexibility: The model should be capable of accurately approximating any probability distribution, including multi-modal ones.

ii) Closed-form likelihood: Models with analytically computable likelihood functions are preferable, avoiding slower, less accurate approximation methods that can lead to noisy gradient estimates during training.

iii) Closed-form sampling: The ability to generate samples analytically, particularly through inversion sampling, is vital. Alternatives like the thinning algorithm are inefficient, less precise, and unsuitable for parallel hardware like GPUs.

These properties ensure the model is versatile, computationally efficient, and capable of generating accurate samples, which is particularly important for network traffic simulation. The LogNormMix TPP~\cite{shchur2019intensity} that we leverage in this work satisfies all three of these properties, whilst other existing approaches are sub-optimal, offering different trade-offs between the three. In this work we propose a multi-mark TPP that bridges ML- and TPP-based approaches to network traffic simulation.
\subsection{Why focus on background traffic?}
Simulating attack traffic is relatively straightforward: red team exercises, replay frameworks, and automated attack tools make it easy to generate adversarial events in controlled environments. In contrast, capturing or synthesizing realistic background traffic—reflecting the diversity, timing, and structure of normal user activity—is far more difficult. This benign traffic forms the baseline that intrusion detection systems and human analysts rely on to distinguish between normal and malicious behavior. In our own deployments, we collect real attack traces from red team events, making background traffic the missing component needed to build realistic and reusable training environments.

Background traffic is also often unavailable in sensitive environments due to legal, privacy, or operational constraints. Even when replay logs exist, they may lack the diversity, timeliness, and contextual nuance required for generalizable IDS testing or cyber range training. Synthetic generation helps address these gaps—offering flexible, reusable traffic traces that preserve structural and temporal realism without relying on privileged data access.

\section{TempoNet Methodology}
\subsection{High-level overview}
\textbf{Network Headers.} Let $D = {(t_i, \textbf{x}_i)}_{i=1}^N$ be a header trace dataset, where:
\begin{itemize}
    \item N is the total number of records (packets or flows)
    \item $t_i \in \mathbb{R}^+$ is the timestamp of the $i^{th}$ record
    \item $\textbf{x}_i \in X$ is the vector of header fields for the $i^{th}$ record,
\end{itemize}
and $X := X_1 \times X_2 \times ... \times X_{m-1}$, where $X_j$ represents the space of possible values for the $j^{th}$ header field (e.g., IP addresses, ports, protocols).  

\begin{figure*}[h]
\includegraphics[width=\textwidth]{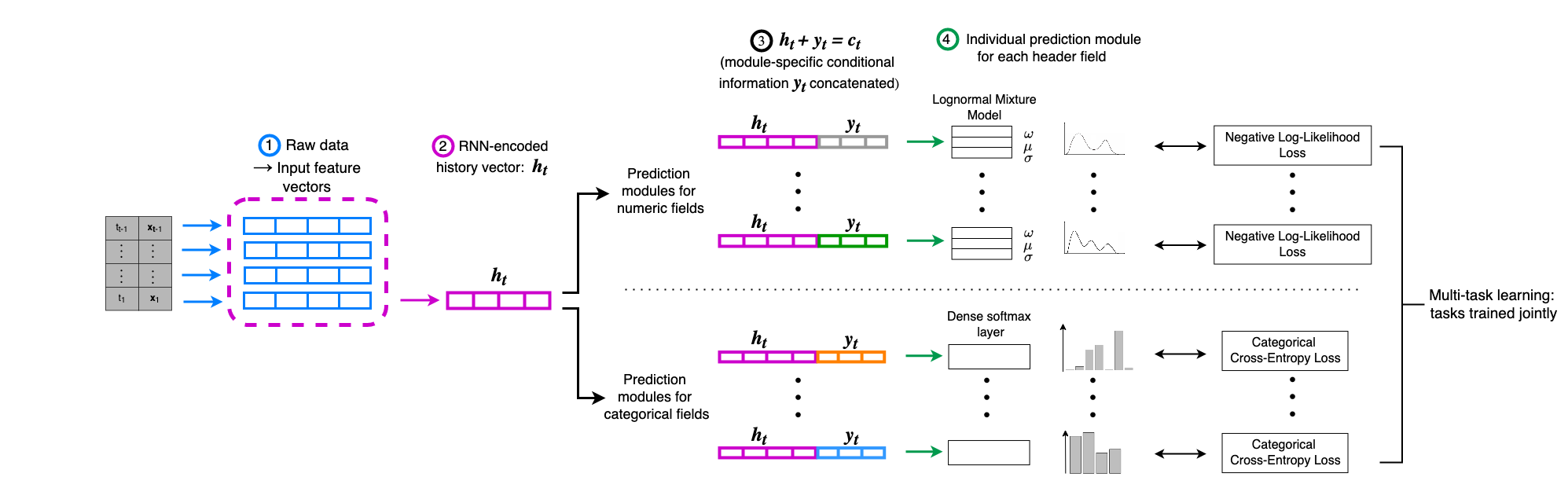}
    \caption{TempoNet architecture.}
    \label{fig:architecture}
\end{figure*}

\textbf{Sequential Dependency and History Encoding.} A key characteristic of network traffic is that the attributes of an incoming packet or flow header can be influenced by the sequence of preceding headers. To capture this historical dependency, we utilize a recurrent neural network (RNN), specifically an LSTM in our implementation. The RNN processes the sequence of past events, denoted as $\mathcal{H}_t:=\{(t_i, \textbf{x}_i) \in D : t_i < t\}$, and condenses this information into a fixed-length vector $\mathbcal{h}_t \in \mathbb{R}^H$. This vector serves as a compact representation of the event history up to time $t$, providing temporal context for subsequent field prediction modules. Conditioning flow generation on this learned history is critical for modeling burstiness, long-range autocorrelation, and diurnal rhythms. Fixed-window or memory-less methods struggle to reproduce such dependencies, whereas RNNs offer a compact yet expressive summary that enables TempoNet to capture both fine-grained bursts and broader seasonal cycles.

\textbf{TempoNet design.} We frame flow and packet header generation as a multi-task machine learning problem. In this approach:
\begin{enumerate}
    \item Each task, representing the generation of a specific header field, has its own prediction module and associated negative log-likelihood (NLL) loss function. 
    \item The fixed-dimensional RNN history vector $\mathbcal{h}_t$  serves as a shared backbone input for all prediction modules.
    \item Each prediction module learns the distribution of its respective header field, conditioned on: i) The history embedding vector $\mathbcal{h}_t$, ii) Other relevant header trace fields, where applicable. Task-specific output layers act as specialized decoders for each prediction module, tailored to the unique characteristics of the corresponding header field.
    \item An equal weighting optimization strategy is used, where the loss weight for each task is set to 1/$m$ in
    every iteration, where $m$ denotes the number of tasks.
\end{enumerate}
\textbf{Multi-task Learning vs. Separate Models.} TempoNet’s multi-task design ensures that each task (e.g., timing, protocol, ports) benefits from a shared temporal context via a single RNN encoder. This avoids the redundancy of training separate models, each with its own RNN, which would inflate parameter counts, increase training time, and risk inconsistent temporal modeling. By consolidating shared structure into one encoder, TempoNet improves parameter efficiency and enforces consistent temporal dependencies across heterogeneous header fields.\\
\textbf{Modeling heterogeneous fields.} To accommodate the diversity of header fields, numeric attributes are modeled with a log-normal mixture model as the prediction module (trained using NLL loss), while categorical attributes are modeled with a categorical distribution (trained with categorical cross-entropy loss). Specifically:
\begin{itemize}
    \item \textit{Inter-arrival times: }flow and packet inter-arrival times are modeled by the log-normal mixture model TPP, known as LogNormMix.~\cite{shchur2019intensity}. 
    \item \textit{Source/destination IP addresses:} Host pair selection is formulated as a source-driven process, where the source IP address is selected first using the categorical distribution for the source module. Destination IP generation is then treated as a multi-class classification problem conditioned on the given source IP.
    \item \textit{Size and flow duration:} These are dependent on the specific host pair~\cite{huang2023datacenter} and are each modeled with a log-normal mixture model, conditioned on the host pair.
    \item \textit{Ports and Protocol:} To capture dependence of ports and protocol on the specific host pair, we model these fields with a categorical distribution, conditioned on the host pair, with the destination ports further conditioned on the protocol. 
\end{itemize}
Due to the high number of tasks for NetFlow generation, we separate NetFlow header generation into 2 stages: 1) Temporal communication dynamics: namely, who is communicating with whom and at what times, which includes flow inter-arrival times, host pair, and flow duration. 2. Communication characteristics: which models the specific attributes of the communication, given the host pair from the first stage. This includes ports, network protocol, and total number of bytes.
\subsection{Neural network architecture}
The high-level architecture of TempoNet is shown in Fig. \ref{fig:architecture}.
The model consists of various prediction modules. Each NetFlow/packet header is generated sequentially, one field at a time, starting from the arrival time and the host IP pair. The generation of each header field can be conditioned on other features (in addition to the history), such as host-pair, by concatenating the history embedding $\mathbcal{h}_{i}$ with a metadata vector $\mathbcal{y}_i$ that encodes the additional features. The resulting context vector $\mathbcal{c}_i = |\mathbcal{h}_i||\mathbcal{y}_i|$ then serves as input into each header field's prediction module. 
Here we outline the composition of each of the prediction modules:
\subsubsection{Log-normal mixture modules}
The LogNormMix TPP~\cite{shchur2019intensity} models the probability distribution of time-stamped event streams using a log-normal mixture model. Specifically, it models the inter-event arrival times, $\{\tau_i\}$, of isolated events by directly learning the conditional probability density function of the time $\tau_i$ until the next event: 
\begin{equation}
\label{lognormal-mixture}
    p(\tau\vert\omega, \mu, \sigma) = \sum_{k=1}^K \omega_k \dfrac{1}{\tau \sigma_k \sqrt{2\pi}} \exp{-\dfrac{(\log \tau - \mu_{k})^2}{2\sigma_k^2}}
\end{equation}
Here $K$ is the number of mixture components, $\omega$ denotes the mixture weights, and $\mu$ and $\sigma$ are the mixture means and standard deviations, respectively. 
These log-normal mixture model distribution parameters are defined to be an affine function of the context vector $\mathbcal{c}_i = |\mathbcal{h}_i||\mathbcal{y}_i|$ via:
\begin{align*}
    \omega_i &= \text{softmax}(V_{\omega}\mathbcal{c}_i + b_{\omega}) \\
    \sigma_i &= \exp(V_{\sigma}\mathbcal{c}_i + b_{
    \sigma
    }) \\
    \mu_i &= V_{\mu}\mathbcal{c}_i + b_{\mu}
\end{align*}
where $\{V_{\omega}, V_{\sigma}, V_{\mu}, b_{\omega}, b_{\sigma}, b_{\mu}\}$ are learnable parameters, and the softmax and exponential transformations enforce essential constraints on the distribution
parameters.\\
\textbf{Rationale for Log-normal Mixtures.} We adopt log-normal mixtures for inter-arrival time modeling due to their universal approximation property on $\mathbb{R}^+$, which allows them to flexibly capture diverse, multi-modal distributions of network timing. Alternative distributions present practical drawbacks for training: log-logistic models exhibit undefined moments, complicating likelihood-based learning, while gamma mixtures require reparameterized sampling, hindering efficient optimization. The log-normal family, by contrast, admits closed-form likelihoods and inversion sampling, enabling stable training and efficient simulation. These properties make it well suited for capturing both the heavy-tailed bursts and regular periodicities characteristic of real network traffic.\\
Existing TPP literature, however, has generally been limited to modeling only one~\cite{shchur2019intensity} or two~\cite{moore2022modelling} categorical marks, with spatio-temporal variants typically restricted to two categorical or continuous marks for spatial coordinates~\cite{chenneural, zhou2022neural}. To the best of our knowledge, TempoNet is the first to augment a TPP model to jointly learn more than 3 fields of mixed numeric and categorical type, enabling richer and more realistic network traffic simulation. \\
In TempoNet, the log-normal mixture is applied not only to inter-arrival times but also to other numeric fields where temporal or host-dependent variability is critical: 

\textit{Timestamps:} The metadata vector $\mathbcal{y}_i$ encodes temporal-related information such as the day of the week and time of day, to help the model learn the seasonality of network traffic on the cyber range's simulated corporate network.

\textit{Size and flow duration:} For both the NetFlow/packet size and the flow duration prediction module, the metadata vector $\mathbcal{y}_i$ incorporates host IP pair embeddings. The size and flow duration distributions are modeled using the same log-normal mixture approach as the inter-event arrival times $\tau$, following equation \ref{lognormal-mixture}.

\subsubsection{Categorical prediction modules}
Categorical prediction modules are used to model selection of the host IP pair, as well as the ports and protocol. 
 
During generation, the class with the highest likelihood is selected.

\textit{Source IP module:} This incorporates a feedforward network and softmax

, generating output logits that define a categorical distribution over network nodes. The implemented source IP feedforward network in our experiments comprises two fully connected layers with an intervening tanh activation layer.

\textit{Destination IP, Port, and Protocol modules:}

The prediction module to learn a categorical distribution for the destination IP, port, and protocol fields consists of a fully connected layer plus softmax. 

\subsection{Event generation with TempoNet}
TempoNet's architecture facilitates straightforward header sampling. Numeric fields like the inter-event time $\tau$, flow/packet size, and flow duration are sampled from the log-normal mixture using standard mixture model sampling:
\begin{enumerate}
    \item Sample the mixture component:
    \vspace{-1mm}
    \begin{equation*}
        \mathbf{z} \sim \textit{Categorical}(\boldsymbol{\omega})
    \end{equation*}
    \vspace{-0.5mm}
    where $\mathbf{z}$ is a one-hot vector of size $K$.

    \item Sample from the unit Normal distribution:
    \vspace{-1mm}
    \begin{equation*}
        \varepsilon \sim \mathcal{N}(0,1)
    \end{equation*}
    \vspace{-2mm}

    \item Transform the unit normal sample $\varepsilon$ to the log-normal distribution for the mixture component $\mathbf{z}$:
    \vspace{-1mm}
    \begin{equation*}
        \tau = \exp(\boldsymbol{\sigma}^{\mathsf{T}}\mathbf{z}\,\varepsilon + \boldsymbol{\mu}^{\mathsf{T}}\mathbf{z})
    \end{equation*}
    \vspace{-2mm}
\end{enumerate}

The host IP pair, ports and protocol of the header are drawn from their respective categorical distributions, with the sampled host IP pair fed as input into the respective selection modules via the metadata vector $\mathbcal{y}_i$ as appropriate. 

\section{Evaluation}
Our evaluation aims to assess whether TempoNet can produce synthetic traffic that is realistic, diverse, novel, and compliant with protocol rules. We benchmark against leading methods from each category (GAN, LLM, Bayesian), evaluating their ability to replicate both static distributions and temporal dependencies critical for security applications.

\subsection{Datasets}
We evaluate on four publicly available datasets that vary in topology, scale, and protocol mix. LANL~\cite{turcotte2019unified} contains internal traffic from a large enterprise network. CIDDS~\cite{ring2017creation} provides flow-level data with labeled attacks and external scanning traffic. DC~\cite{benson2010network} is a lightweight campus network capture. IoT~\cite{alsaedi2020ton_iot} uses the ToN\_IoT \emph{network-flow} subset from a heterogeneous IoT/IIoT testbed across multiple sites. 
We include this dataset primarily as a robustness stress test under distribution shift, since its lab-generated nature and temporal inconsistency make it less directly comparable to enterprise or campus traces. Full dataset descriptions and preprocessing steps are in Appendix~\ref{appendix:datasets}.

\subsection{Baselines}
We compare TempoNet to the following state-of-the-art GAN, LLM, and Bayesian Network generation models\footnote{We initially intended to include the auto-regressive model STAN in our experiments, but like the authors of \cite{schoen2024tale}, we were unable to reproduce it and unfortunately had to rule it out.}.
\begin{enumerate}
    \item NetShare~\cite{yin2022practical} is a GAN-based approach that adapts and extends the DoppelGANger~\cite{lin2019generating} framework to address the specific challenges associated with generating synthetic IP header traces. It improves fidelity by capturing intra- and inter-epoch correlations, and enhances scalability through fine-tuning and parallel training.  However, NetShare is known to be computationally expensive, with the original study requiring a cluster of 200 CPU cores and 1.92~TB of RAM.  
    
    \item GReaT\cite{borisov2022language} generates synthetic tabular data using transformer-based neural networks. The GReaT approach involves two main stages: fine-tuning a pretrained LLM on the tabular data and sampling from the fine-tuned model to generate synthetic data.
    \item The BN$_{bins}$ model studied in \cite{schoen2024tale} is a Bayesian Network-based generative approach that uses a quantile-based discretization strategy for numerical features. This method captures inter-feature dependencies through Conditional Probability Tables (CPTs), enabling efficient modeling of the joint distribution of network flow data. We use the \textit{bnlearn} Python library implementation of BN$_{bins}$. 
\end{enumerate}    

Other approaches, such as the VAE-based method Encore~\cite{huang2023datacenter}, focuses solely on modeling flow sizes between host pairs, and is unable to model the other critical header fields (such timestamps or ports) or network-wide dynamics. Similarly, the diffusion-based model in \cite{jiang2023generative} faces challenges with long training times, slow inference, and limited scalability, having only been tested on a small dataset of 30,000 records. As a result both approaches are unsuitable for high-fidelity, scalable network traffic modeling, and were ruled out.

\subsection{Experiments and Metrics}\label{sec:metrics}

We simulate 9-10 weeks of packet/NetFlow header data for  of the 3 main datasets and evaluate the generated traffic against both the real training and test sets using a comprehensive set of metrics. Our evaluation blends established statistical tests with domain-specific diagnostics inspired by real-world security needs such as cyber range exercises.\\
\textbf{Reproducibility considerations.} For NetShare, we trained using 10 CPU cores and 1.5~TB of RAM on the LANL and DC datasets, compared to the 200 cores and 1.92~TB recommended in the original paper. While this resource gap may have impacted performance, prior independent evaluations~\cite{schoen2024tale} also report instability and limited fidelity even under the recommended setup, suggesting that these challenges are partly architectural. We therefore present NetShare as a strong but resource-intensive baseline, while emphasizing TempoNet’s efficiency advantages.\\
\textbf{Realism, Diversity, Novelty, and Compliance.}
We adopt the framework introduced by~\cite{schoen2024tale}:
\begin{itemize}
    \item \textbf{Realism} measures alignment with real data distributions. This includes marginal distribution metrics like Jensen-Shannon Divergence (JSD) and Earth Mover’s Distance (EMD), as well as inter-feature metrics like Pairwise Conditional Distribution (PCD) for numerical correlations and Contingency Matrix Differences (CMD) for categorical dependencies. Lower scores indicate better fidelity. 

     \item \textbf{Diversity} is evaluated using the coverage metric, which assess how broadly synthetic data spans the real distribution's support. Coverage quantifies the number of synthetic neighborhood spheres that contain each real sample. A low score means that many real samples do not have corresponding generated synthetic samples nearby, revealing that the synthetic distribution fails to sufficiently represent the variability present in the real distribution.

     \item \textbf{Novelty} is assessed via the Membership Disclosure (MD) metric. MD measures the degree to which synthetic samples differ from the real training data by computing the proportion of synthetic samples that are unusually close based on Hamming distance—to real training data. Rather than rewarding lower MD, we compare against the test set MD score, indicating that the synthetic data has a similar degree of novelty and overlap as real, unseen traffic. Extremely low MD may indicate underfitting or insufficient coverage of realistic behaviors.

    \item \textbf{Compliance} is measured by Domain Knowledge Check, which validates conditions such as: (i) packet sizes within valid ranges (e.g., minimum 40 bytes for TCP, 28 bytes for UDP) and (ii) appropriate relationships between port numbers and protocols (e.g., port 80 for HTTP, port 53 for DNS). A detailed description of the compliance check is provided in Appendix C.
\end{itemize}
We use the original TALE metric implementations to ensure consistent and replicable evaluation.

\begin{table*}[t]
\centering
\setlength{\tabcolsep}{2.8pt} 
\renewcommand{\arraystretch}{1.05} 
\scriptsize 
\caption{Comparison of 4 models across metrics for realism, diversity, novelty, and compliance. 
The "Real test set" row provides a baseline for unseen real data. Bold \textbf{black} text marks 
the best result, and bold \textcolor{gray}{\textbf{gray}} text marks the second best. 
(↑↑) indicates higher is better, (↓↓) lower is better, and (=) closer to the test set is better.}
\label{tab:tale-metrics}
\vspace{-1mm} 
\resizebox{0.96\textwidth}{!}{%
\begin{tabular}{|l|l|c|c|c|c|c|c|c|c|}
\hline
\multicolumn{2}{|c|}{} & \multicolumn{3}{c|}{\textbf{Realism}} & \textbf{Diversity} & \multicolumn{2}{c|}{\textbf{Realism \& Diversity}} & \textbf{Novelty} & \textbf{Compliance} \\ \hline
\textbf{Dataset} & \textbf{Model} & CMD (\(\downarrow\)) & PCD (\(\downarrow\)) & Density (\(\uparrow\)) & Coverage (\(\uparrow\)) & EMD (\(\downarrow\)) & JSD (\(\downarrow\)) & MD (=) & DKC (\(\downarrow\)) \\ \hline
\multirow{5}{*}{\textbf{LANL}} 
& {\ul Real test set} & {\ul 0.049} & {\ul 0.009} & {\ul 0.918} & {\ul 0.802} & {\ul 0.0011} & {\ul 0.11} & {\ul 7.065} & {\ul 3e-5} \\ 
& NetShare & 0.327 & 0.471 & 0.384 & 0.269 & \textcolor{gray}{\textbf{0.0349}} & 0.35 & 5.374 & 0.050 \\
& GReaT & 0.053 & \textcolor{gray}{\textbf{0.006}} & \textcolor{gray}{\textbf{0.809}} & \textcolor{gray}{\textbf{0.68}} & 0.124 & 0.06 & \textbf{6.819} & \textbf{2e-4} \\
& BN$_{bins}$ & \textbf{0.004} & 0.2312 &  0.552 & 0.288 & 0.043 & \textbf{0.01} & 5.938 & 1.4e-3 \\
& TempoNet (Ours) & \textcolor{gray}{\textbf{0.043}} & \textbf{0.005} & \textbf{0.820} & \textbf{0.767} & \textbf{0.0006} & \textcolor{gray}{\textbf{0.03}} & \textcolor{gray}{\textbf{6.007}} & \textcolor{gray}{\textbf{8e-4}} \\ \hline
\multirow{4}{*}{\textbf{CIDDS}} 
& {\ul Real test set} & {\ul 0.068} & {\ul 0.0098} & {\ul 0.915} & {\ul 0.840} & {\ul 0.00008} & {\ul 0.12} & {\ul 7.36} & {\ul 5e-4} \\ 
& GReaT & \textcolor{gray}{\textbf{0.078}} & \textcolor{gray}{\textbf{0.0038}} & \textbf{1.053} & \textbf{0.837} & {\color{gray}\textbf{0.0001}} & \textcolor{gray}{\textbf{0.09}} & \textbf{6.46} & \textbf{8e-4} \\
& BN$_{bins}$ & 0.125 & 0.491 & 0.557 & 0.494 & 0.06 & 0.14 & 6.25 & \textcolor{gray}{\textbf{0.008}} \\
& TempoNet (Ours)& \textbf{0.043} & \textbf{0.0018} & \textcolor{gray}{\textbf{0.684}} & \textcolor{gray}{\textbf{0.784}} & \textbf{0.00005} & \textbf{0.05} & \textcolor{gray}{\textbf{6.37}} & 0.023 \\ \hline
\multirow{5}{*}{\textbf{DC}} 
& {\ul Real test set} & {\ul 0.177} & {\ul 0.073} & {\ul 0.578} & {\ul 0.389} & {\ul 0.022} & {\ul 0.38} & {\ul 7.956} & {\ul 0.006} \\
& NetShare & 0.250 & 0.234 & 0.301 & \textcolor{gray}{\textbf{0.230}} & 0.072 & 0.72 & 3.009 & \textcolor{gray}{\textbf{6e-6}} \\
& GReaT & 0.205 & 0.232 & 0.303 & 0.223 & {\color{gray}\textbf{0.060}} & 0.34 & 6.819 & \textbf{1e-6} \\
& BN$_{bins}$ & \textbf{0.070} & \textbf{0.030} & \textcolor{gray}{\textbf{0.328}} & 0.092 & {\color{black}0.24} & \textbf{0.14} & \textcolor{gray}{\textbf{6.986}} & 0.008 \\
& TempoNet (Ours) & \textcolor{gray}{\textbf{0.074}} & \textcolor{gray}{\textbf{0.120}} & \textbf{0.420} & \textbf{0.321} & \textcolor{black}{\textbf{0.036}} & \textcolor{gray}{\textbf{0.20}} & \textbf{7.050} & 0.008 \\ \hline
\multirow{4}{*}{\textbf{IoT}} 
& {\ul Real test set} & {\ul {\color{black}0.688}} & {\ul {\color{black}0.865}} & {\ul {\color{black}0.94}} & {\ul {\color{black}0.1357}} & {\ul {\color{black}0.0002}} & {\ul {\color{black}0.551}} & {\ul {\color{black}5.83}} & {\ul {\color{black}0.022}} \\ 
& GReaT &{\color{gray}\textbf{0.115}} & \textcolor{gray}{\textbf{0.88}} & {\color{black}\textbf{0.63}} & {\color{black}\textbf{0.530}} & {\color{black}338} & \textcolor{black}{\textbf{0.092}} & \textcolor{gray}{\textbf{6.62}} & {\color{black}0.043} \\
& BN$_{bins}$ & {\color{black}\textbf{0.114}} & {\color{black}\textbf{0.451}} &{\color{black} 0.12} & {\color{black}0.19} & {\color{gray}\textbf{20.00}} & {\color{gray}\textbf{0.101}} & \textcolor{black}{7.21} & \textcolor{gray}{\textbf{0.039}} \\
& TempoNet (Ours)& \textcolor{black}{0.20} & \textcolor{gray}{\textbf{0.88}} & \textcolor{gray}{\textbf{0.48}} & \textcolor{gray}{\textbf{0.31}} & \textcolor{black}{\textbf{0.00019}} &\textcolor{black}{0.180} & \textcolor{black}{\textbf{5.86}} & \textcolor{black}{\textbf{0.037}} \\ \hline
\end{tabular}
\hspace{0.001cm} 
\begin{tabular}{|c|}
\hline
\textbf{Rank} \\ \hline
Average (\(\downarrow\))\\ \hline
- \\ 
\textcolor{black}{3.75} \\
\textcolor{gray}{\textbf{2.25}} \\
\textcolor{black}{2.5} \\
\textcolor{black}{\textbf{1.5}} \\ \hline
- \\ 
\textcolor{black}{\textbf{1.5}} \\
2.875 \\
\textcolor{gray}{\textbf{1.63}} \\ 
\hline
- \\ 
\textcolor{black}{3.25} \\
\textcolor{black}{2.75} \\
\textcolor{gray}{\textbf{2.25}} \\
\textcolor{black}{\textbf{1.63}}\\ \hline
- \\ 
\textcolor{black}{\textbf{1.875}} \\
\textcolor{black}{2.125} \\
\textcolor{black}{\textbf{1.875}} \\ \hline

\end{tabular}}
\vspace{-2mm}
\end{table*}

\textbf{Categorical Field Fidelity.}
We further assess per-field realism using JSD on key categorical variables: source IP, destination IP, destination port, and protocol. These metrics evaluate whether the model captures structural properties like service use and communication asymmetries.

\textbf{Temporal and Higher-Order Fidelity.}
To reflect real-world use in cyber ranges, we evaluate the following.
\begin{itemize}
    \item \textbf{Temporal Fidelity} is assessed using EMD on inter-arrival times, flow durations, and seasonality features (hour-of-day, day-of-week). We visualize how well each model reproduces real-world inter-arrival times using quantile–quantile (Q–Q) plots  and use boxplots to illustrate alignment in daily and weekly seasonality patterns.

        \item \textbf{Host IP pair-level Fidelity} evaluations are carried out to assess host IP selection realism. To do this, we examined the 30 most active source/destination IP pairs from the training set for each dataset, and visualized the proportion of packets/flows sent by each source IP and received by each destination IP using box plots.
\end{itemize}
\textbf{Interpreting test set performance.}
TempoNet and all baselines are trained on the training split of each dataset. The Real test set scores reported in our tables are not intended as performance targets, but serve as calibration baselines that illustrate the natural variability between two real-world samples from the same distribution. This contextualization is especially important in metrics like EMD and Diversity, where theoretical optima (e.g., EMD = 0, Diversity = 1) may be unattainable in practice. We include test set values to provide a sanity check: if even two real samples diverge on a metric, then perfect synthetic alignment is neither expected nor required. For metrics like MD, the test set also provides a meaningful novelty baseline. This framing helps calibrate expectations and interpret synthetic performance more realistically.

\section{Results}

\subsection{Overall realism, diversity, novelty and compliance.}\label{Sec:Realism}
We evaluate model performance across the four criteria using the metrics from Section \ref{sec:metrics}, with results shown in Table \ref{tab:tale-metrics}. TempoNet performs strongly across all metrics, ranking first on LANL and DC datasets and second on CIDDS, highlighting its robustness across diverse datasets. 

\textbf{Realism.} TempoNet achieves the best density scores on LANL and DC, reflecting its ability to learn and reproduce the joint data distribution. Its design---explicitly conditioning predictions on relevant header fields (e.g., ports on Host IP pairs)---enhances this alignment. TempoNet also excels on CMD and PCD, which measure attribute correlations, consistently ranking first or second. GReaT LLM also performs well, coming in second to TempoNet on both LANL and CIDDS. Meanwhile, BN$_{bins}$ leads realism on the DC dataset. NetShare, as noted in prior work \cite{schoen2024tale}, performs poorly overall on realism.

\textbf{Diversity.} 
TempoNet shows strong coverage across datasets, generating samples that reflect the breadth of the real data. GReaT ranks second, particularly strong on CIDDS. Examining EMD and JSD reveals further nuances: TempoNet and BN$_{bins}$ better capture categorical features (higher JSD), while GReaT and NetShare are relatively stronger on numerical ones. TempoNet consistently balances diversity and realism most effectively.

\textbf{Novelty.} We use Membership Disclosure (MD) to assess overfitting. Rather than treating lower MD as inherently better, since real-world network traffic naturally includes repeated structures, we benchmark model MD scores against the Real test set (MD between 7–8 across datasets). An MD score close to the test set indicates that a model captures natural variability without overfitting. GReaT mathces this best on LANL and CIDDS, while TempoNet leads on DC. NetShare, with much lower MD scores, generates traffic overly dissimilar from real patterns. Overall, TempoNet and GReaT best preserve novelty without sacrificing fidelity.

\textbf{Compliance.} 
Domain Knowledge Compliance (DKC) assesses whether generated traffic adheres to network protocol rules. Notably, the Real test sets themselves exhibit non-zero DKC scores (e.g., 0.006 for DC), indicating that a small proportion of NetFlows and packets in the original datasets do not fully conform to domain constraints—likely due to real-world anomalies or noise. On the DC dataset, TempoNet and BN$_{\text{bins}}$ closely mirror this compliance profile, suggesting they faithfully replicate both typical behavior and occasional irregularities present in the real data. GReaT achieves the lowest DKC scores overall, reflecting strong protocol adherence. Across all datasets, most models produce DKC values comparable to or lower than the test set. Exceptions include NetShare on LANL and TempoNet on CIDDS, which show slightly higher DKC values. Nevertheless, TempoNet maintains strong overall compliance, balancing realism with adherence to domain-specific rules. An expanded discussion of the violated DKC rules, their causes, and implications is provided in Appendix C.

\textbf{IoT Robustness Stress Test.} 
Finally, to probe robustness under distribution shift, we evaluated models on the ToN\_IoT dataset. 
Unlike the enterprise and campus traces, IoT is lab-generated, temporally inconsistent, and exhibits extreme skew (with a handful of devices dominating traffic). As expected, no model achieves strong absolute scores on this dataset. The Real test set itself diverges substantially from nominal ideals (e.g., CMD, PCD, coverage, JSD), reflecting mismatches between training and test distributions. This suggests that enforcing those ideals on IoT would amount to overfitting to the training set. In this light, TempoNet’s results, while not always the closest to the ideals, are better 
aligned with the Real test set’s deviations, whereas competing models achieve 
superficially lower errors that reflect overfitting rather than fidelity. 
Notably, TempoNet and the GReaT LLM baseline tie for best overall performance, 
underscoring the difficulty of this stress-test setting.

\textbf{Overall Interpretation.} LLM-based models like GReaT excel in compliance and numerical feature modeling, while TPP-based models like TempoNet offer superior realism, diversity, and generalization. BN$_\text{bins}$ performs well on correlation-based metrics like CMD but struggles with density and coverage, indicating limitations in capturing the full data distribution. Overall, NetShare under-performs across all criteria, showing limited realism, diversity, and compliance.

\subsection{Per-Field Categorical Fidelity.}\label{sec:categorical-fidelity}
Table~\ref{tab:field-jsd} reports the Jensen–Shannon Divergence (JSD) between real and generated distributions for key categorical fields: source IP, destination IP, destination port, and protocol. These metrics assess how well each model captures structural properties of network traffic. TempoNet achieves consistently low JSD scores, demonstrating strong alignment with real-world feature distributions.

\begin{table}[]
    \centering
    \setlength{\tabcolsep}{4pt} 
    \renewcommand{\arraystretch}{1.2} 
    \small 
\caption{Per-field JSD for categorical headers, reported as mean (std) over 10 trials.}
    \label{tab:field-jsd}
    \resizebox{\columnwidth}{!}{\begin{tabular}{|l|l|c|c|c|c|}
        \hline
        \textbf{Dataset} & \textbf{Model} & \textbf{Source IP} & \textbf{Dst IP} & \textbf{Dst Port} & \textbf{Protocol}\\
        & & \textbf{JSD} (\(\downarrow\)) & \textbf{JSD} (\(\downarrow\))& \textbf{JSD} (\(\downarrow\)) &\textbf{JSD}  (\(\downarrow\))\\ \hline
        \multirow{4}{*}{\textbf{LANL}} 
        & NetShare & 0.490 (.01) & 1.49 (.44) & 0.45 (.04)& 0.125 (.023)\\
        & GReaT & 0.150 (6e-4) &  0.11 (8e-4) & \textbf{0.01 (.001)} & \textbf{0.004 (.001)} \\
        & BN$_{\text{bins}}$ & \textcolor{gray}{\textbf{0.127(4e-4)}} & \textbf{0.033 (.001)} & 0.02 (.001) & 0.088 (.061)\\
        & TempoNet & \textbf{0.044 (.001)}  & \textcolor{gray}{\textbf{0.058 (.001)}} & \textcolor{gray}{\textbf{0.02 (.003)}} & \textcolor{gray}{\textbf{0.005 (.002)}}\\ 
        \hline
        \multirow{3}{*}{\textbf{CIDDS}} 
        & GReaT & \textcolor{gray}{\textbf{0.32 (3e-4)}} & \textcolor{gray}{\textbf{0.66 (2e-4)}}& \textcolor{gray}{\textbf{0.06 (4e-4)}}& \textbf{0.003 (5e-4)} \\
        & BN$_{\text{bins}}$ & \textcolor{gray}{\textbf{0.32 (9e-5)}} & \textcolor{gray}{\textbf{0.66 (3e-4)}} &\textcolor{gray}{\textbf{0.06 (.001)}}  & 0.016 (.001)\\
        & TempoNet & \textbf{0.05 (.004)}  & \textbf{0.08 (8e-4)} & \textbf{0.04 (8e-4)} & \textcolor{gray}{\textbf{0.005 (.001)}}\\ \hline
        
        \multirow{4}{*}{\textbf{DC}} 
        & NetShare & 0.83 (.00) & 0.83 (.00) & 0.82 (3e-3) & 0.44 (.06)\\
        & GReaT &  0.23 (8e-4) &  0.41 (2e-4) & 0.82 (3e-5) &\textbf{0.061(6e-4)}  \\
        & BN$_{\text{bins}}$ & \textbf{0.19 (5e-4)} & \textcolor{gray}{\textbf{0.40 (2e-4)}} & \textbf{0.28 (1e-4)} & \textcolor{gray}{\textbf{0.087 (3e-4)}}\\
        & TempoNet & \textcolor{gray}{\textbf{0.20 (.001)}}  & \textbf{0.27 (1e-3)} & \textcolor{gray}{\textbf{0.29 (8e-4)}} & 0.106 (1e-3) \\ \hline
        \multirow{3}{*}{\textbf{IoT}}
        & GReaT & \textbf{0.075 (.0008)} & \textcolor{gray}{\textbf{ 0.098 (.0006)}}& \textbf{0.12 (0.0006)}& \textbf{0.079 (0.0006)} \\
        & BN$_{\text{bins}}$ & \textcolor{gray}{\textbf{0.11 (0.0005)}} & \textbf{0.022 (0.0006)}&\textcolor{gray}{\textbf{0.20 (0.0009)}}  & \textcolor{gray}{\textbf{0.087 (0.001)}}\\
        & TempoNet & 0.28 (0.07)  & 0.24 (0.08) & \textcolor{gray}{\textbf{0.20 (0.02)}} & 0.10 (0.005)\\ \hline
    \end{tabular}}
\end{table}

\subsection{Temporal fidelity.}\label{Sec:temporal}
\begin{table}[t]
    \centering
    \setlength{\tabcolsep}{4pt} 
    \renewcommand{\arraystretch}{1.2} 
    \small
    \caption{EMD between real and simulated data, reported as mean (std) over 10 trials.}
    \label{tab:temp-generation-results}
    \resizebox{\columnwidth}{!}{\begin{tabular}{|l|l|c|c|c|c|}
        \hline
        \textbf{Dataset} & \textbf{Model} & \textbf{Time Deltas} & \textbf{Hourly} & \textbf{Daily} & \textbf{Duration}\\
        & & \textbf{EMD} (\(\downarrow\)) & \textbf{EMD} (\(\downarrow\))& \textbf{EMD} (\(\downarrow\)) &\textbf{EMD}  (\(\downarrow\))\\ \hline
        \multirow{4}{*}{\textbf{LANL}} 
        & NetShare & 30.54 (8.28) &\textcolor{gray}{\textbf{ 0.34 (0.01)}} & 0.47 (0.03)& \textcolor{gray}{\textbf{133.7 (56.9)}}\\
        & GReaT & 6.39 (0.02) & 0.48 (0.01) & 0.23 (0.01) & 441 (50.3)\\
        & BN$_{\text{bins}}$ & \textcolor{gray}{\textbf{3.93 (0.03)}} & 0.36 (0.01) & \textcolor{gray}{\textbf{0.18 (0.01)}}& 163886 (1083)\\
        & TempoNet & \textbf{0.58 (0.10)} & \textbf{0.15 (0.06)} & \textbf{0.10 (0.01)} &\textbf{0.66 (.01)}\\ 
        \hline
        \multirow{3}{*}{\textbf{CIDDS}} 
        & GReaT & \textcolor{gray}{\textbf{3e-4 (3e-6)}} & 2.48 (0.03) & \textbf{0.57 (0.02)} & \textcolor{gray}{\textbf{0.13 (4e-4)}}\\
        & BN$_{\text{bins}}$ & 1.70 (1e-3) & \textcolor{gray}{\textbf{2.41 (4e-3)}} & \textcolor{gray}{\textbf{0.61 (1e-3)}} &1.16 (2e-4)\\
        & TempoNet & \textbf{1e-4 (2e-5)} & \textbf{0.41 (0.08)} & 0.83 (0.05) & \textbf{0.02 (.002)}\\ \hline
        
        \multirow{4}{*}{\textbf{DC}} 
        & NetShare & 6e-4 (8e-5) & 0.93 (0.8) & 0.96 (1.0) &-\\
        & GReaT & \textcolor{black}{\textbf{6.5e-6 (1e-6)}} & \textcolor{gray}{\textbf{0.41 (7e-3)}} & \textcolor{gray}{\textbf{0.12 (7e-3)}} &-\\
        & BN$_{\text{bins}}$ & 0.01 (5e-5) & \textbf{0.29 (0.01)} & 0.28 (2e-3) &-\\
        & TempoNet & \textcolor{gray}{\textbf{7e-5 (8e-6)}} & 0.52 (0.09) & \textbf{0.09 (0.03)}& -\\ \hline
         \multirow{3}{*}{{\color{black}\textbf{IoT}}} 
        & {\color{black}GReaT} & {\color{black}2962530 (9368342)} & {\color{black}\textbf{0.40 (.08)}} & {\color{gray}\textbf{0.48 (.01)}} & {\color{black}\textbf{9.43 (.04)}}\\
        & {\color{black}BN$_{\text{bins}}$} & {\color{gray}\textbf{105170.5 (90518.0)}} & {\color{gray}\textbf{0.43 (0.27)}} & {\color{black}0.84 (0.51)} & {\color{black}9248.7 (36.4)}\\
        & {\color{black}TempoNet} & {\color{black}\textbf{0.34 (0.17)}} & {\color{black}0.62 (0.05)} & {\color{black}\textbf{0.13 (0.02)}} & {\color{gray}\textbf{37.33 (11.46)}}\\ \hline
    \end{tabular}}
\end{table}

We assess the temporal realism and diversity of the models using Earth Mover’s Distance (EMD) on four key features: inter-arrival times (time deltas), hour of day, day of week, and flow duration, across LANL, CIDDS, and DC datasets. We further visualize temporal fidelity via Q–Q plots and seasonality boxplots. Additional visualizations, including CIDDS Q–Q plots and LANL/DC seasonality trends, are included in Appendix B.

\textbf{Quantitative results.} Table \ref{tab:temp-generation-results} shows that TempoNet consistently achieves the lowest EMD scores across most metrics and datasets, indicating strong alignment with real distributions, particularly for inter-arrival times (Time Deltas), flow durations, and broader periodic trends (Hour of Day and Day of Week). GReaT performs well in some cases (e.g., day-of-week on CIDDS) but is inconsistent. BN$_{\text{bins}}$ shows moderate performance, especially on hour-of-day trends, while NetShare yields the highest EMD values, reflecting weak temporal realism.

\textbf{Q--Q plots of inter-arrival times.} Figure~\ref{fig:QQ-deltas-LANL} (with CIDDS shown in Appendix B (Figure \ref{fig:QQ-deltas-CIDDS})) presents Q--Q plots comparing real and generated inter-arrival time quantiles. Points lying near the red diagonal indicate close distributional match; systematic deviations reflect under- or overestimation of inter-arrival times at specific quantiles. The black points are empirical quantile pairs between the real and synthetic data distributions.

\begin{figure}[h]
    \centering
\includegraphics[width=6cm]{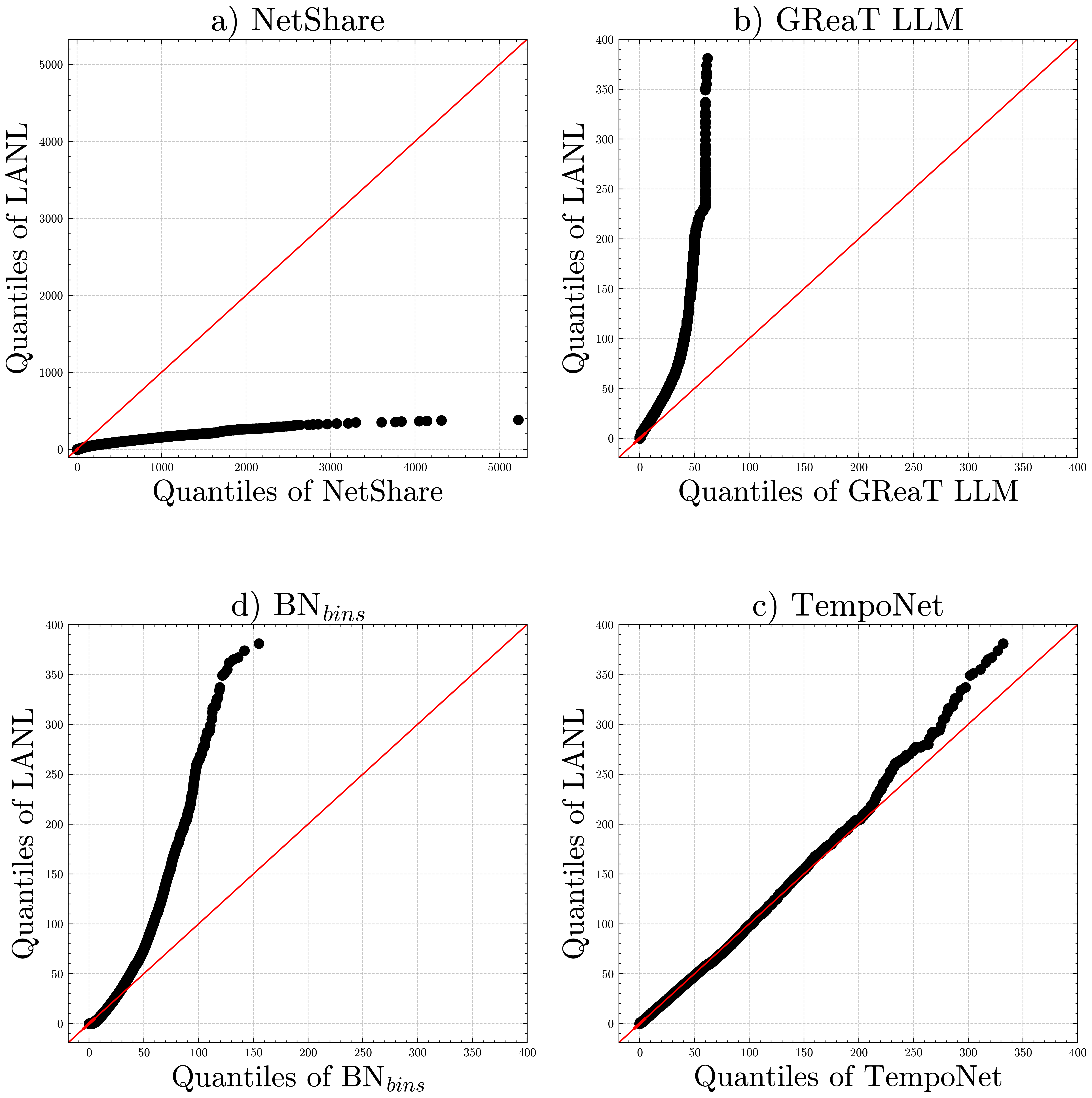}
    \caption{Q--Q plots of LANL data flow inter-arrival times: generated vs ground truth.}
    \label{fig:QQ-deltas-LANL}
\end{figure}

TempoNet closely tracks the diagonal across datasets, indicating strong fidelity. NetShare diverges sharply at higher quantiles, producing overly long delays. GReaT performs better at lower quantiles but underestimates at the tail. BN$_{\text{bins}}$ shows dataset-dependent artifacts, including step-like quantization on CIDDS. Overall, TempoNet offers the most realistic inter-arrival time modeling.

\textbf{Boxplots of daily seasonality.} 

\begin{figure}[h]
\includegraphics[width=\columnwidth]{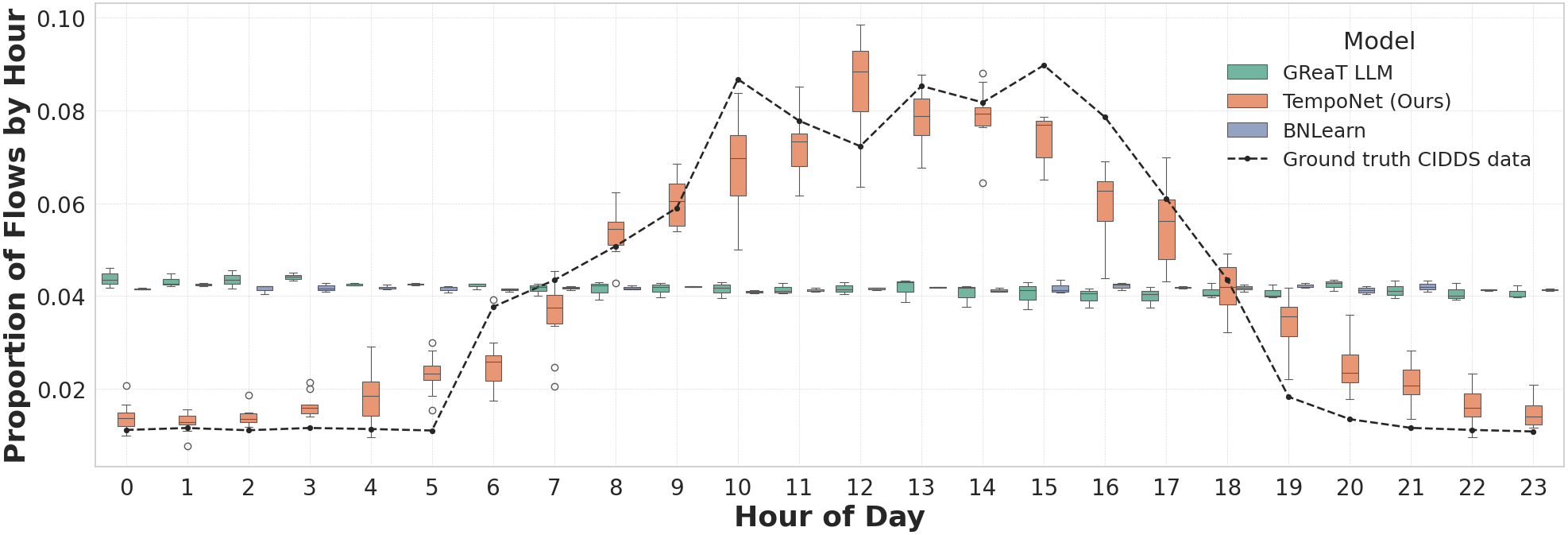}
    \caption{Daily seasonality - CIDDS dataset.}
    \label{fig:CIDDS-hour-hist}
\end{figure}

Figure~\ref{fig:CIDDS-hour-hist} shows hour-of-day trends for CIDDS (LANL in Appendix B (Figure~\ref{fig:LANL-hour-hist})). TempoNet is the only model to closely reproduce the daily rhythms--e.g., mid-day peaks and overnight lulls, which others generate flat or noisy distributions.
This suggests that TempoNet captures periodicity at both coarse and fine scales.

\textbf{Boxplots of weekly seasonality.}
 Weekly trends are shown in Figure~\ref{fig:LANL-day-hist} for LANL (DC in Appendix B (Figure~\ref{fig:DC-weekday-hist})). TempoNet again most closely mirrors the ground truth, while BN$_{\text{bins}}$ captures rough trends with some variance.  GReaT and NetShare fail to match the pattern, producing flatter or distorted outputs. 

\begin{figure}[h]
\includegraphics[width=\columnwidth]{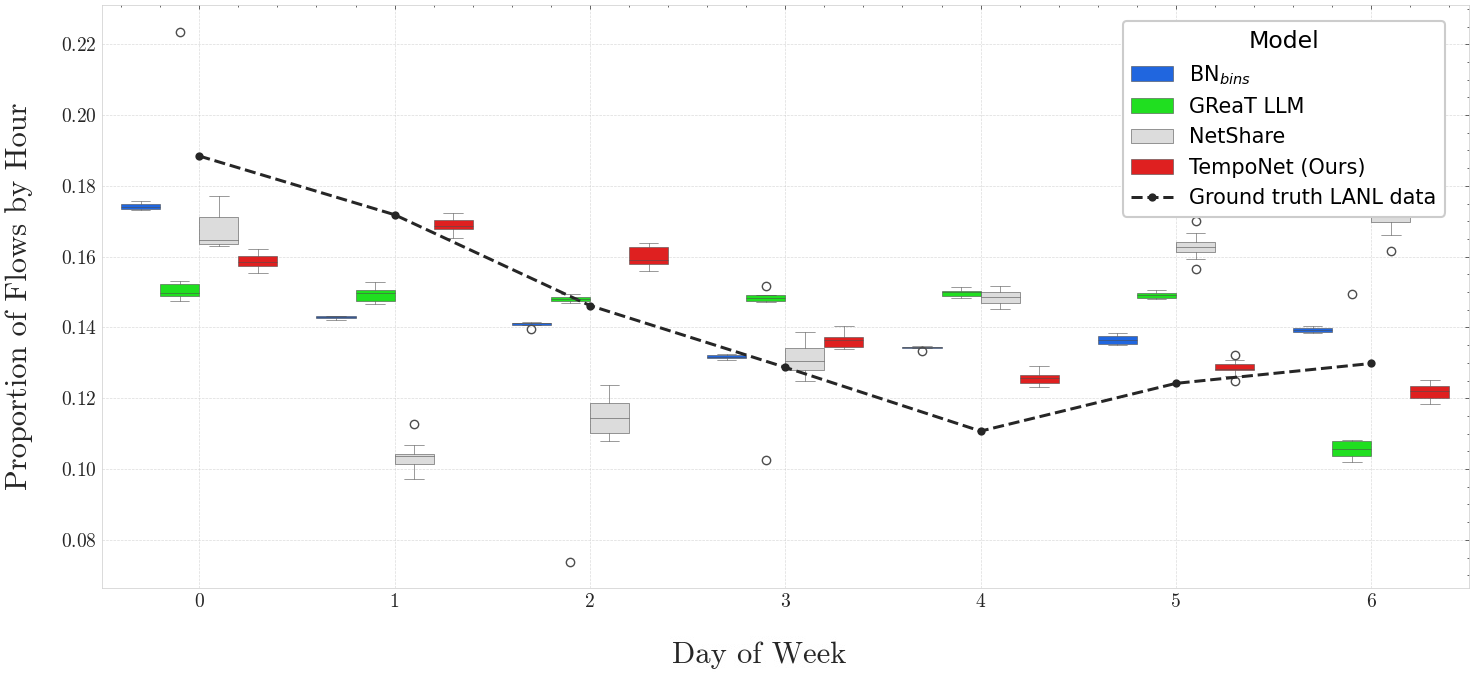}
    \caption{Weekly seasonality - LANL dataset.}
    \label{fig:LANL-day-hist}
\end{figure}

\textbf{Summary.}
Across datasets and metrics, TempoNet demonstrates the strongest temporal fidelity, replicating both micro-level timing and macro-level seasonal patterns. Other models lag due to inconsistent or oversimplified generation of temporal features.

\subsection{Host IP pair-level fidelity.}\label{sec:pair}
We evaluate the realism of host IP selection by comparing the distribution of packets or flows across the top 30 most active source/destination IP pairs in each dataset. Figure~\ref{fig:LANL-pair-hist} shows results for LANL; results for CIDDS and DC are in Appendix B (Figures~\ref{fig:CIDDS-pair-hist} and~\ref{fig:DC-pair-hist}).

\begin{figure}[h]
\includegraphics[width=\columnwidth]{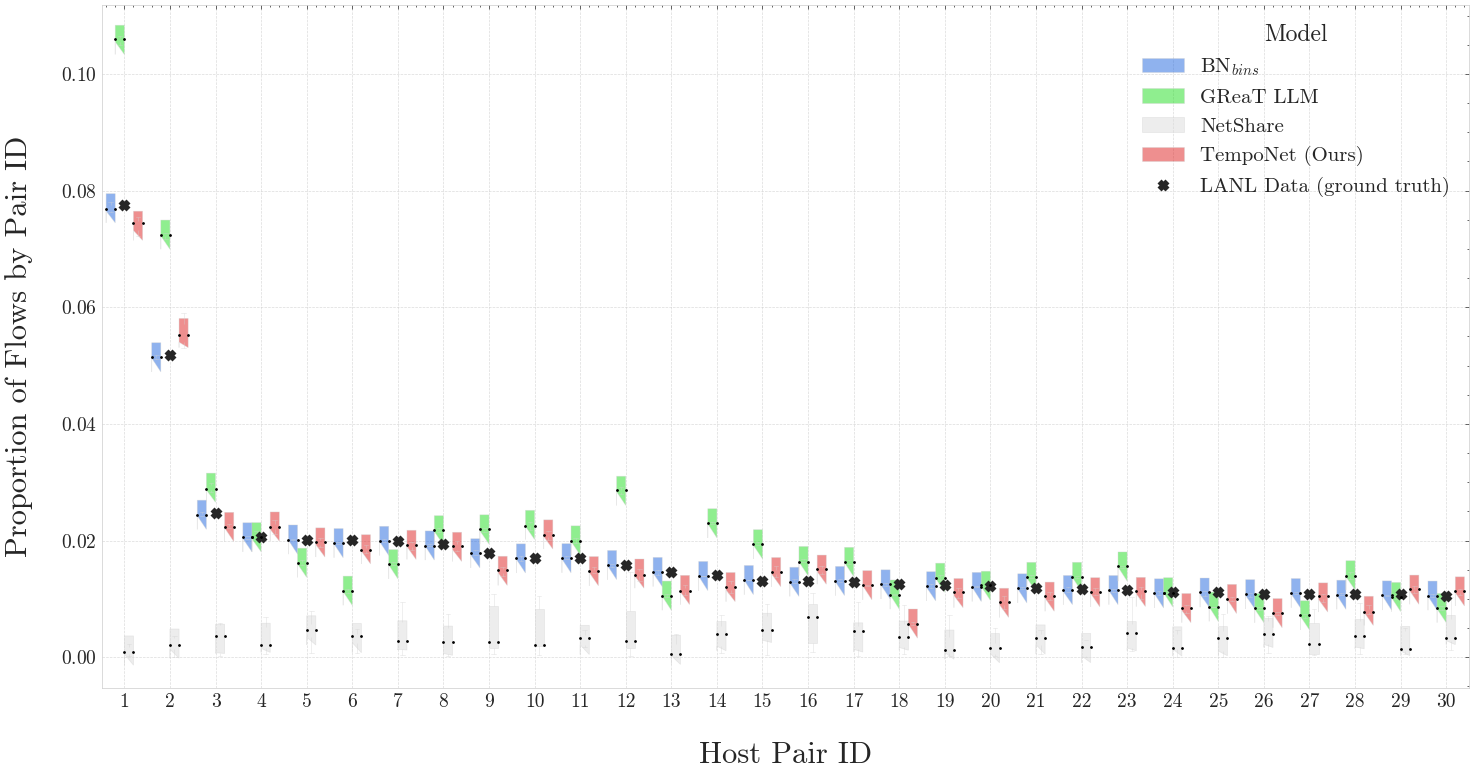}
    \caption{Host Pair Analysis - LANL dataset.}
    \label{fig:LANL-pair-hist}
\end{figure}

For the LANL Dataset, TempoNet and BN$_{\text{bins}}$ closely match the real distribution, capturing the relative proportions across host pairs. GReaT moderately aligns but overestimates the 2 most active host pairs. NetShare underestimates across the board, failing to reflect the skew in real traffic.

For CIDDS (Figure \ref{fig:CIDDS-pair-hist} in Appendix B), TempoNet again aligns well, especially for high-activity pairs. BN$_{\text{bins}}$ underestimates top pairs but preserves general trends. GReaT overestimates the dominant pairs and shows less accurate overall modeling.

For DC (Figure \ref{fig:DC-pair-hist} in Appendix B), TempoNet, BN$_{\text{bins}}$ and GReaT track the overall shape but exhibit localized deviations. NetShare performs poorly, producing flat, unrealistic distributions.

\textbf{Summary.} TempoNet consistently delivers the best host pair fidelity, maintaining both global trends and fine-grained variations across datasets. BN$_{\text{bins}}$ performs well on the LANL dataset but struggles with the other 2 datasets with larger host spaces. GReaT ranks third overall, while 
NetShare exhibits the weakest alignment.

\subsection{Training Time and Model Size Comparison}

Table~\ref{tab:training-times} summarizes the training times and model sizes (measured by number of trainable parameters) for each method. NetShare required substantially more memory (1.5~TB RAM) and parallel CPU resources (10 cores), reflecting the high cost of training GANs on complex tabular data. Moreover, NetShare’s publicly available implementation does not support GPU acceleration, further contributing to its extended runtime.

In contrast, all other models—including TempoNet—were trained using a single NVIDIA H100 GPU (96GB VRAM), with 1 CPU core and 100GB system memory. Among them, GReaT required the longest training time and had the largest model size, with over 80 million parameters. Despite its scale, GReaT did not consistently outperform other models, highlighting the cost-performance trade-off in large LLM-based generators.

TempoNet, by comparison, is two orders of magnitude smaller than GReaT, yet achieved comparable or superior results across realism, diversity, and security utility metrics. This highlights the architectural efficiency of combining multi-task learning with a structured temporal model. BN$_{\text{bins}}$, while extremely compact and fast to train, underperformed on metrics requiring temporal fidelity. These results demonstrate that TempoNet strikes a compelling balance between modeling capacity, speed, and scalability.
\begin{table}[t]
    \centering
    \setlength{\tabcolsep}{4pt}      
    \renewcommand{\arraystretch}{0.98} 
    \scriptsize                      
    \vspace{-1.5mm}                 
    \caption{Model training time and size.}
    \label{tab:training-times}
    \begin{tabular}{llcc}
        \toprule
        \textbf{Dataset} & \textbf{Model} & \textbf{Train Time (min)} & \textbf{\# Params} \\
        \midrule
        \multirow{4}{*}{\textbf{LANL}} 
        & NetShare & $>$3{,}000 & 179{,}369 \\
        & GReaT & 42.14 & 81{,}912{,}576\\
        & BN$_{\text{bins}}$ & 0.02 & 4{,}652\\
        & TempoNet & 4.90 & 41{,}976\\
        \midrule
        \multirow{3}{*}{\textbf{CIDDS}} 
        & GReaT & 146.00 & 81{,}912{,}576\\
        & BN$_{\text{bins}}$ & 0.08 & 13{,}518 \\
        & TempoNet & 29.7  & 68{,}629\\
        \midrule
        \multirow{4}{*}{\textbf{DC}} 
        & NetShare & $>$3{,}000 & 179{,}369\\
        & GReaT & 195.0 & 81{,}912{,}576\\
        & BN$_{\text{bins}}$ & 0.04 & 135{,}059 \\
        & TempoNet & 29.7  & 128{,}494  \\
        \bottomrule
    \end{tabular}
    \vspace{-2.5mm}                 
\end{table}

\subsection{Impact on Intrusion Detection Performance}
While generating attack traffic is relatively straightforward---via red teaming in staging environments, replaying previously captured exploits, or automating attacks using known toolchains, the greater challenge lies in simulating the complex, dynamic behavior of real users—the benign background traffic. Without this, intrusion detection systems (IDS) risk being trained on unnaturally clean data, and cyber range exercises can lose operational realism, making detection trivially easy due to the lack of natural network noise.

To assess whether TempoNet addresses this gap, we evaluated its utility in supporting anomaly-based intrusion detection. This is not a benchmark of IDS performance, but a test of whether synthetic benign traffic preserves enough realism to support effective security tasks. We conducted a three-stage evaluation using the CIDDS-001 dataset:

\textbf{(1) Synthetic Benign Traffic Generation:} We trained TempoNet on the benign portion of the CIDDS-001 dataset and generated synthetic background traffic, previously evaluated in Sections~\ref{Sec:Realism} and~\ref{Sec:temporal}.

\textbf{(2) Integration of Real Attacks:} We overlaid real, labeled attack flows from the CIDDS-001 test set onto the synthetic benign traffic, preserving original timestamps to maintain realistic temporal context. These attacks are used solely for evaluation and are never seen during training.

\textbf{(3) Evaluation on Real Data:} 
We evaluate anomaly detectors trained \emph{only on benign traffic} and tested on a held-out week of CIDDS-001 containing both benign and attack flows, to avoid data leakage and mirror deployment conditions. For each training source (real, TempoNet-generated, GReaT-generated), we fit: {\color{black} i) an \emph{Isolation Forest} on point-wise flow features; (ii) a \emph{sequence LSTM autoencoder} over fixed-length windows \(L\), scoring by mean squared reconstruction error; and (iii) a \emph{One-Class SVM} (RBF). For all three, the decision threshold/hyperparameters are chosen on a benign-only validation split, then held fixed for testing (eg. for the LSTM Autoencoder we fixed the anomaly threshold at the 92nd percentile of reconstruction errors). We report class-specific F1 (benign/attack) and overall accuracy.} 

{\color{black} \textbf{Results.}
Table~\ref{tab:ids-results} shows that TempoNet largely preserves detector behavior. For Isolation Forest and the LSTM Autoencoder, TempoNet tracks Real closely (all metric deltas \(\le 0.03\)). For the One-Class SVM, TempoNet shows a small but consistent drop ($\approx$\(-0.01\) Benign-F1, \(-0.04\) Attack-F1, \(-0.02\) Accuracy), yet remains far closer to Real than the GReaT LLM background.

By contrast, the GReaT LLM background exhibits reduced benign fidelity: it inflates One-Class SVM performance while markedly degrading LSTM Autoencoder attack detection (requiring a substantially looser threshold to recover near-real scores), indicating distortions of higher-order/temporal statistics that change class separability.

\textbf{Takeaway.} TempoNet is task-faithful: swapping Real\(\rightarrow\)TempoNet leaves IDS decisions essentially unchanged for tree and sequence models and only modestly shifts a kernel detector, whereas the LLM background induces non-faithful, detector-dependent shifts (inflations or degradations).
}

\begin{table}[ht]
\centering

\caption{IDS evaluation on CIDDS dataset.}
\label{tab:ids-results}
\begin{tabular}{p{1.7cm}lccc}
\toprule
\textbf{IDS} & \textbf{Metric} & \textbf{Real} & \textbf{TempoNet} & \textbf{GReaT LLM} \\
\midrule
\multirow{3}{=}{\makecell{Isolation\\Forest}} 
 & Benign F1 & 0.95 & 0.95 & 0.86 \\
 & Attack F1 & 0.73 & 0.70 & 0.72 \\
 & Accuracy  & 0.92 & 0.91 & 0.82 \\
\midrule
\multirow{3}{=}{\makecell{{\color{black}LSTM}\\{\color{black}Autoencoder}}} 
 & {\color{black}Benign F1 }& {\color{black}0.97} & {\color{black}0.98}
 & {\color{black}0.95}\\
 & {\color{black}Attack F1} & {\color{black}0.84} & {\color{black}0.83} & {\color{black}0.50} \\
 & {\color{black}Accuracy}  & {\color{black}0.95} &{\color{black}0.96} &  {\color{black}0.91} \\
 \midrule
\multirow{3}{=}{\makecell{{\color{black}One-Class}\\{\color{black}SVM}}} 
 & {\color{black}Benign F1 }& {\color{black} 0.970} & {\color{black} 0.960}
 & {\color{black} 0.992}\\
 & {\color{black}Attack F1} & {\color{black} 0.845} & {\color{black} 0.803} & {\color{black} 0.955} \\
 & {\color{black}Accuracy}  & {\color{black} 0.950} &{\color{black} 0.933} &  {\color{black} 0.987} \\
\bottomrule
\end{tabular}
\end{table}

\section{Discussion \& Future Directions}

In this work, we introduced TempoNet, a temporal point process (TPP)-based generative model for network traffic simulation. TempoNet demonstrated strong performance across key metrics, including realism, diversity, novelty, and compliance, outperforming state-of-the-art GAN, LLM, and Bayesian Network-based approaches in many areas. While other models capture certain temporal trends, TempoNet is the only approach that consistently models both fine-grained and high-level seasonality across datasets. This fidelity to temporal patterns is essential for applications requiring realistic and dynamic network simulations, such as cyber range training, cyber deception, and networking tasks like intrusion and anomaly detection.

\subsection{Discussion}
\textbf{Temporal Fidelity for Security and Simulation:} 
TempoNet's ability to replicate temporal seasonality provides significant advantages across both security and networking applications. In intrusion detection systems (IDS), particularly those based on anomaly detection, the accuracy of benign traffic modeling has a direct impact on false positive rates and overall detection reliability. Much of the existing IDS benchmarking literature focuses on injecting attack traffic into synthetic or anonymized traces, often overlooking the importance of realistic background dynamics. This can lead to overfitting and misleading performance evaluations.

TempoNet addresses this gap by capturing complex temporal patterns—such as daily and weekly seasonality—that reflect real-world usage. In operational training scenarios, particularly in cyber ranges, accurate simulation of benign traffic is essential for both realistic defender experience and reliable performance testing of defence mechanisms. Likewise, in cyber deception systems, realistic temporal dynamics help create engaging environments that sustain adversary interaction and reduce detectability. Beyond security-specific applications, TempoNet-generated datasets also have the potential to benefit broader networking tasks, including telemetry analysis, predictive maintenance, and benchmarking network performance under realistic, time-varying conditions.

\textbf{Forecasting and Prediction:} A unique strength of TPP-based models is their ability to forecast event sequences. While our current focus is on simulation, TempoNet's architecture can naturally extend to predictive tasks. Its log-normal mixture model allows not only the generation of future traffic patterns but also the estimation of their likelihood, making TempoNet a valuable tool for proactive tasks such as capacity planning, anomaly detection, and predictive maintenance in network management.

\textbf{Comparison with Existing Methods:} TempoNet addresses gaps in existing methods by integrating TPPs with multi-task learning. GAN-based models, such as NetShare, show strength in reproducing numerical marginal feature distributions but struggle with temporal dependencies. LLM-based models, like GReaT, handle categorical features effectively but lack explicit temporal modeling. Bayesian Networks (e.g., BN$_{\text{bins}}$) capture static correlations but fail to replicate dynamic patterns. 

TempoNet uniquely combines fine-grained temporal modeling with broader seasonal patterns while preserving relationships among header fields, positioning it as a superior solution for diverse use cases. Its use of a flexible, multi-modal log-normal mixture distribution enables rich sampling across the learned data manifold, eliminating the need for explicit diversity-enhancement mechanisms. The architecture is trained to model the overall data distribution rather than memorize individual instances, which naturally promotes generalization and sample-level diversity. This is supported by our Coverage and Density metrics, which quantify how broadly the generated data spans the support of the real distribution.

Regarding novelty, our use of the Membership Disclosure (MD) metric confirms that TempoNet avoids direct memorization. While MD is typically used as a privacy metric, it also serves as a proxy for sample uniqueness, helping to demonstrate that synthetic data points deviate sufficiently from the training set. Together, these results suggest that TempoNet achieves both high fidelity and meaningful novelty, without requiring additional mechanisms for diversity control.

\subsection{Limitations and Future Work}

{\color{black}\textbf{Limitations:} Despite its strengths, TempoNet has several limitations. First, while it captures daily and weekly rhythms well, sparse events remain challenging. The log-normal mixture, being smooth, has a tendency to either overgenerate around rare events or omit them entirely. Second, capturing multiple seasonalities (hourly, daily, weekly) remains difficult—metadata helps with individual cycles, but not with their interactions. Third, long-range dependencies such as rare bursts or gradual drifts are hard to model across multiple time scales. Fourth, the model synthesizes only packet- and flow-header fields rather than
payloads, which narrows applicability for payload-centric tasks (e.g., deep packet inspection). Fifth, our evaluation
focuses on anomaly-based IDS scenarios, leaving open questions about how well TempoNet-generated traffic supports other IDS paradigms (e.g., signature-based or hybrid systems). Finally, dataset recency remains a limitation: modern,
long-duration traces with multi-week coverage are scarce, though we selected datasets with sufficient structural and
temporal diversity—including an IoT dataset—for current benchmarks. These constraints highlight the importance of caution when generalizing results, particularly for edge cases or environments underrepresented in our datasets.}

{\color{black}\textbf{Future Work:}  Addressing these limitations creates several promising directions. Short-term priorities include refining the mixture component to better capture rare bursts
and long-term drifts, alongside formal privacy evaluations (e.g., differential privacy guarantees, membership inference testing) to ensure robust protection when releasing synthetic traces. We also aim to validate TempoNet in operational settings by training IDS, stress-testing security tools, and enhancing cyber range training outcomes. Longer-term, expanding to multi-modal data (e.g., payloads, application-layer metadata) will enable more comprehensive simulations, while exploring TempoNet for traffic prediction could support early anomaly detection, resource optimization, and proactive defense. Finally, hybrid architectures that integrate TempoNet’s temporal modeling with large language models (LLMs) may combine field-level semantic fidelity with explicit temporal dynamics, advancing both simulation and forecasting in network security.}

\section{Related Work}
\textbf{TPPs for network traffic event modeling.}
Temporal Point Processes (TPPs) have been widely applied to capture temporal dynamics in network traffic. BURSE~\cite{6782285} and Moore et al.\cite{moore2016analysis} used Markov-modulated Poisson and Hawkes Processes, respectively, to reproduce burstiness and detect changes in flow. Saha et al.'s NTPP\cite{saha2019learning} combined RMTPP with learning-to-rank, modeling self-excitation and host contention. Price-Williams and Heard~\cite{price2020nonparametric} proposed a Wold process outperforming Hawkes for inter-arrival time modeling.

These methods model fine-grained timing but typically focus on a small number of features (e.g., timestamps), limiting realism for broader simulation tasks. Beyond network traffic, TPPs have also been applied to social media~\cite{zipkin2016point,shchur2019intensity} and spatio-temporal event modeling~\cite{chenneural,zhou2022neural}, but existing models rarely scale beyond two marks. Our work extends this line by modeling 7–8 header fields per event, combining temporal fidelity with full-header realism—marking the first such high-fidelity multi-attribute TPP.

\textbf{ML-based network traffic simulation.} 
ML-based traffic simulation.
GAN-based DoppelGANger~\cite{lin2019generating} and its successor NetShare~\cite{yin2022practical} led early efforts in synthetic header trace generation. {\color{black}NetShare extends DoppelGANger to improve fidelity and privacy, but remains unable to model inter-arrival times and is highly resource-intensive (e.g., requiring 200 CPU cores in the original study), creating reproducibility barriers.} These issues highlight the trade-off between realism and accessibility in GAN-based methods.

Autoregressive models like STAN~\cite{xu2021stan} use CNNs and mixture density networks for multivariate timeseries, but largely focus on intra-host patterns and lack inter-host dynamics. Diffusion models~\cite{jiang2023generative,sivaroopan2024netdiffus} show promise but are limited by slow inference, making them impractical for high-throughput simulation tasks such as cyber range scenarios.

Encore~\cite{huang2023datacenter} employs VAEs with GRUs to model flow sizes between host pairs but does not generalize to other attributes or capture full-network dynamics. In contrast, our method offers an efficient, temporally aware alternative that scales to rich, full-header representations needed for high-fidelity security applications.

\section{Conclusion}
This paper presented TempoNet, a novel approach to network traffic simulation that combines temporal point processes with multi-task learning to generate high-fidelity packet-header and flow-header traces. Our evaluation on real-world NetFlow and pcap datasets demonstrates TempoNet's ability to model all header fields while preserving intricate temporal relationships, significantly outperforming existing methods. The log-normal mixture model TPP component proves particularly effective at capturing complex temporal patterns, including daily and weekly variations, that previous approaches struggled to represent. Beyond its strong performance across realism, diversity, and compliance metrics, TempoNet's ability to generate authentic background traffic addresses a critical need in cybersecurity training and testing environments. These capabilities position TempoNet as a valuable tool for creating more realistic cyber ranges, enabling better training outcomes and more reliable system evaluations.

Importantly, many existing public benchmark datasets used in IDS research contain simulated or anonymized traffic and rely on simplistic background distributions. These datasets limit the generalizability of detection models and under-represent the operational complexity faced by security analysts. Our work addresses this gap by providing a foundation for realistic, temporally coherent background traffic that can be used to train and evaluate intrusion detection models under realistic network load conditions. Future evaluations with TempoNet-enhanced datasets will better reflect the challenges of detecting threats amidst complex, time-varying benign traffic—providing a more credible basis for security model assessment.
\vspace{-1mm}
\section*{Acknowledgment}
\vspace{-1mm}
This work has been supported by the Cyber Security Research Centre Limited, whose activities are partially funded by the Australian Government’s Cooperative Research Centres Programme.
\vspace{-1mm}

\bibliographystyle{IEEEtran}
\bibliography{base}

\appendix
\subsection*{A. Dataset Details}
\label{appendix:datasets}
For our study, our focus on capturing long-term temporal dynamics necessitated the use of datasets spanning at least 3 weeks. As a result, several commonly used datasets were excluded due to the limited timeframe of data captured, which precluded the analysis of important temporal effects in network traffic patterns. Previous works ensure manageable dataset sizes through the use of custom-made synthetic datasets~\cite{xu2021stan}, by capping the number of records of real-world datasets (effectively restricting their duration)~\cite{xu2021stan, yin2022practical,schoen2024tale, huang2023datacenter,jiang2023generative}, or by restricting the number of hosts~\cite{xu2021stan, huang2023datacenter}. To maintain a manageable dataset size while preserving the extended timeframe, we restrict the dataset to a subset of hosts, as detailed for each dataset below. This methodology allowed us to balance the need for a comprehensive temporal perspective with practical constraints on data volume. By doing so, we ensured our dataset captured the full spectrum of temporal variations, including daily and weekly patterns, which are crucial for developing high-fidelity network traffic generation models.\par

{\color{black}\textbf{Required timespan for training data.}
 While our study uses 3-week spans for three of the dataset, this choice was intentional: it enables evaluation of how well models capture long-term temporal patterns, such as workday vs. weekend activity and diurnal cycles. However, TempoNet does not require a full week of data to function effectively. In settings where fine-grained temporal seasonality is less critical, such as evaluating static intrusion detection rules or testing network infrastructure readiness, models can be trained on smaller samples, including traces as short as a single day or hour, provided they capture the key node-level communication structure.

The primary tradeoff of using less data is a reduction in temporal coverage and structural diversity. Rare services and behaviors, as well as weekly usage rhythms, may be under-represented, which can limit the generality of the generated traffic for downstream tasks. Nevertheless, TempoNet remains applicable across a wide range of dataset sizes, allowing deployment in environments with limited historical data availability.}

\textbf{Dataset Composition.} 
Here is the metadata description for the datasets: \newline 
\noindent\textbf{(NetFlow-1) LANL:} The Unified Host and Network Dataset ~\cite{turcotte2019unified} is a subset of network and computer (host) events collected from the Los Alamos National Laboratory enterprise network. We take 4 weeks worth of communications between a subset of 52 nodes, amounting to 202,841 NetFlow records. {\color{black} LANL has one dominant, highly active node (12.6\% of flows), followed by a broad middle tier where 75\% of nodes each contribute between 1–5.5\% of flows; only 4\% of source IPs send less than 0.1\% of flows.}\\
\textbf{(NetFlow-2) CIDDS:} CIDDS~\cite{ring2017creation,ring2017flow} emulates a small business network environment, featuring multiple clients and servers (e.g., email, web) with scripted benign user activities as well as real-world external traffic. The dataset creators incorporate injected malicious traffic to create comprehensive datasets for intrusion detection evaluation, however we sample only from the benign data. We took a subset of 140 most active nodes and sampled 10\% of their communications over 4 weeks of data collection, amounting to 655,359 NetFlow records. {\color{black} CIDDS has one dominant, highly active node (15.3\% of flows), a substantial middle tier where 29\% of IPs each contribute 1–7.5\% of traffic, and a pronounced long tail with 60\% of IPs each sending less than 0.1\%.}\\

{\color{black} \textbf{(NetFlow-3) IoT-TON (IoT): } We use the ToN\_IoT \emph{network-flow} subset~\cite{alsaedi2020ton_iot}, collected from a heterogeneous 
IoT/IIoT testbed across multiple sites. The dataset is lab-generated rather than drawn from continuous 
enterprise traffic, and exhibits irregular temporal consistency (many hours contain no flows, while others 
exceed 6K). After preprocessing we obtain 92 hours of data, amounting to 255,045 flows. 

A key complication is that the train and test subsets differ markedly in their marginal distributions, 
producing atypical “Real” baseline scores in Table~\ref{tab:tale-metrics}. This mismatch limits the 
dataset’s use for precise model benchmarking, but makes it valuable as a robustness \emph{stress test}. 
Indeed, traffic distribution is highly polarized: 78\% of IPs each contribute less than 0.1\% of flows, 
while a handful of devices dominate up to 40\% individually. We therefore retain IoT primarily to probe 
model behavior under distribution shift and extreme skew, rather than to serve as a direct analogue of 
enterprise traces.}

\textbf{(PCAP) Data Center (DC): } The DC dataset utilzes the packet capture "UN1" from the data center studied in the IMC 2010 measurement paper~\cite{benson2010network}. We utilize a subnetwork of 192 most active nodes to obtain 779,364 pcap records. {\color{black} DC has three highly active nodes (9.5\%, 10.5\%, and 12.4\% of flows), a moderate middle tier where 12.5\% of nodes each contribute 1–4.5\%, and about 12\% of IPs send less than 0.1\% of traffic.}\\

\begin{figure}[t]
    \centering
    
    \begin{subfigure}[t]{0.48\textwidth}
        \centering
        \includegraphics[width=\textwidth]{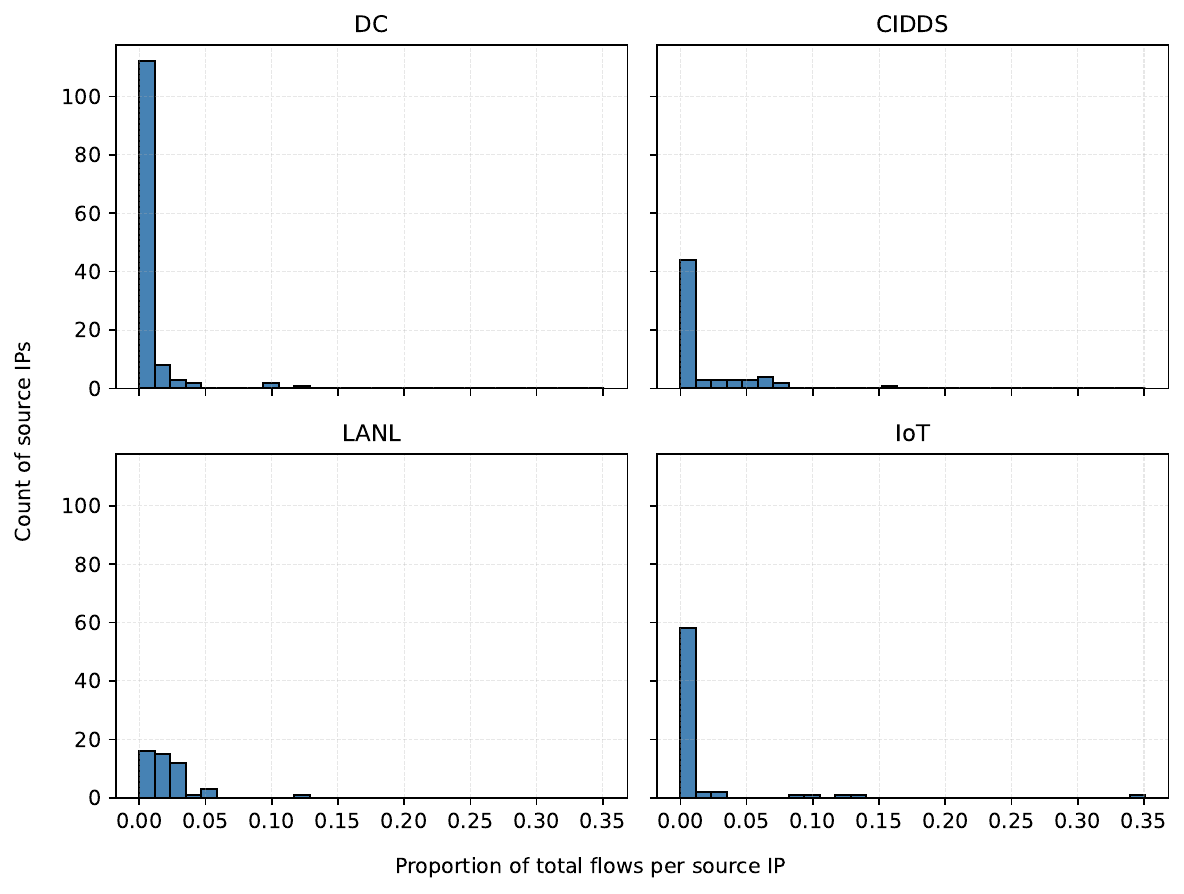}
        \caption{Histograms of per-source IP flow shares across datasets.}
        \label{fig:hist}
    \end{subfigure}
    \hfill
        \begin{subfigure}[t]{0.48\textwidth}
        \centering
        \includegraphics[width=\textwidth]{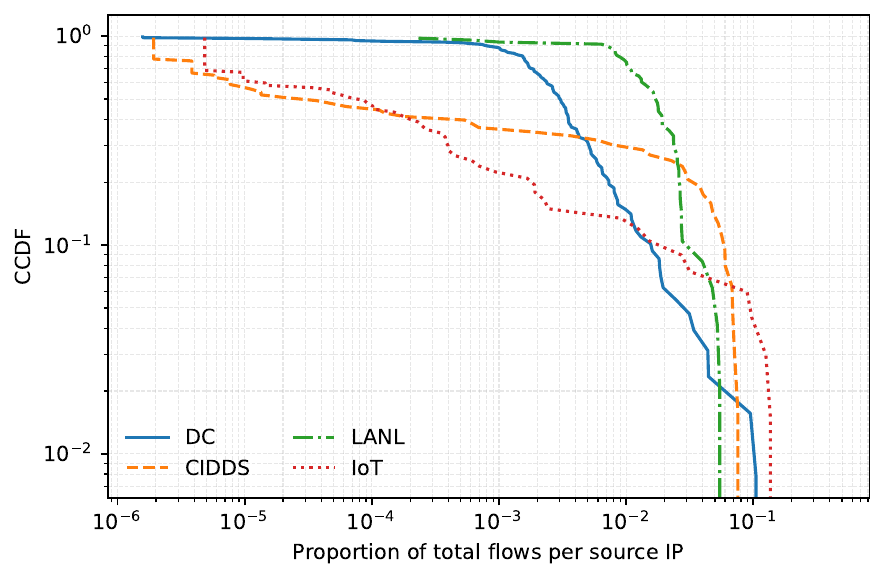}
        \caption{Complementary cumulative distribution function (CCDF) of per-source IP flow shares.}
        \label{fig:cdf}
    \end{subfigure}
    \caption{{\color{black}Activity skew across datasets. The CCDF (left) highlights heavy-tailed behavior, while the histograms (right) illustrate the distribution of flow shares across IPs. Together, these plots show that LANL, CIDDS, DC, and IoT-Ton each exhibit distinct skew profiles (see text for details).}}
    \label{fig:skew}
\end{figure}

{\color{black} Taken together, these four datasets provide a spectrum of network dynamics: LANL with one elephant and a broad middle tier, DC with three elephants and a moderate tail, CIDDS with one elephant and an overwhelming population of mice, and IoT-Ton with extreme polarization between a handful of super-elephants and a large majority of tiny contributors. This diversity ensures that TempoNet is evaluated across both enterprise and IoT settings, under balanced, long-tail, and highly polarized traffic conditions. The corresponding activity histograms and complementary cumulative distribution functions (CCDFs) are shown in Figure~\ref{fig:skew}.}

\subsection*{B. Additional Visualizations and Analyses}
\label{appendix:visuals}
This appendix includes additional visualizations referenced in Sections~\ref{Sec:temporal} and~\ref{sec:pair}, which could not be included in the main paper due to space constraints. These include Q--Q plots, seasonality boxplots (daily and weekly), and host IP pair distributions for datasets not shown in the main text. Together, these plots offer a more complete view of the generative models’ temporal and structural fidelity.

\begin{figure}[htbp!]
\includegraphics[width=\columnwidth]{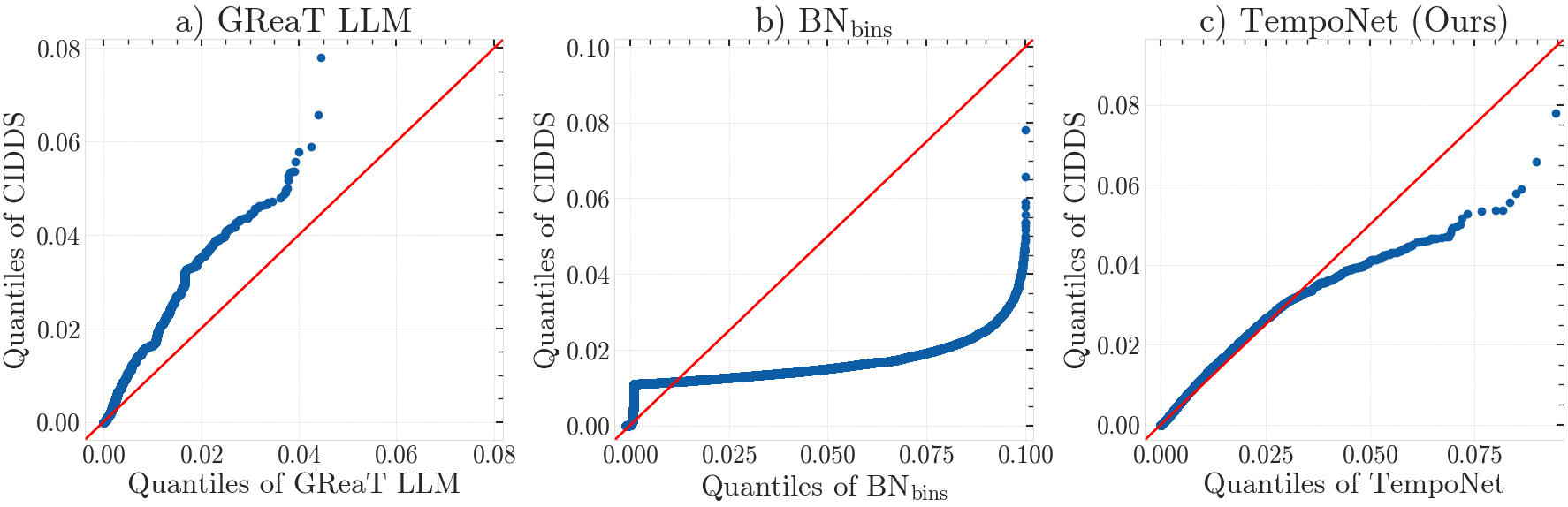}
    \caption{Q--Q Plots of CIDDS Data Flow Inter-arrival Times: Generated Vs Ground Truth.}
    \label{fig:QQ-deltas-CIDDS}
\end{figure}

\begin{figure}[htbp!]
\includegraphics[width=\columnwidth]{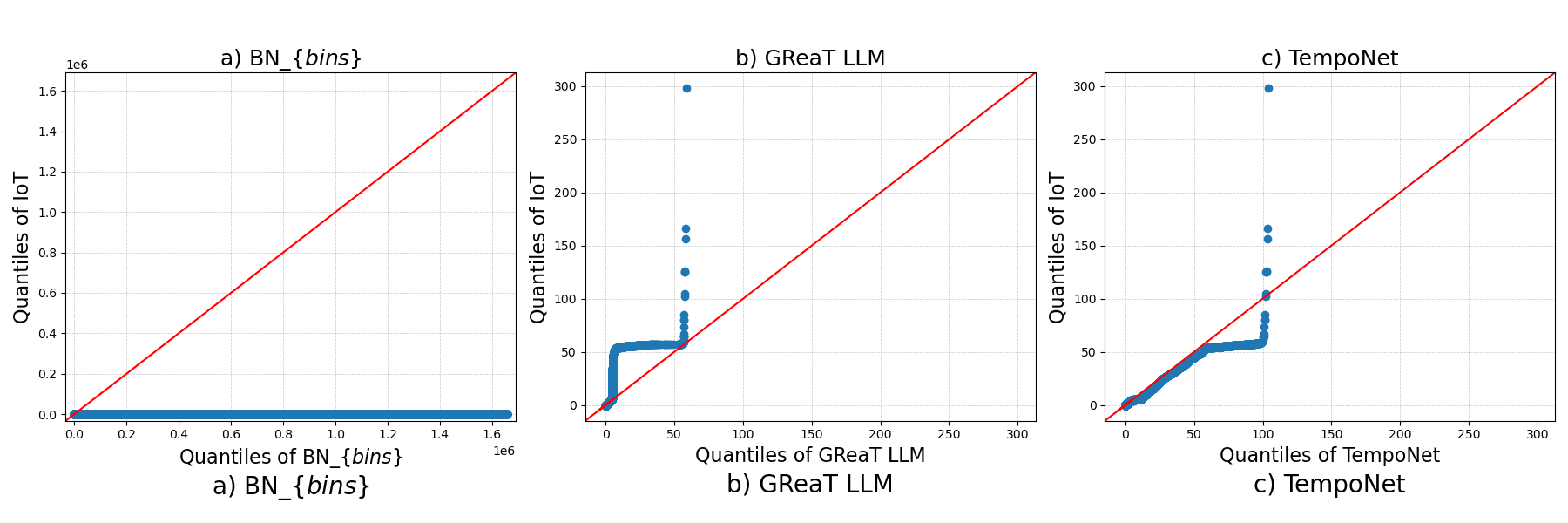}
    \caption{Q--Q Plots of IOT Data Flow Inter-arrival Times: Generated Vs Ground Truth.}
    \label{fig:QQ-deltas-iot}
\end{figure}

\textbf{1. Temporal Fidelity and Failure Cases.}
\textcolor{black}{Figures~\ref{fig:QQ-deltas-CIDDS} and \ref{fig:QQ-deltas-iot} present Q–Q plots of flow inter-arrival times for CIDDS and IoT datasets, respectively. These visualizations highlight where generative models diverge from real temporal dynamics.}

\textcolor{black}{For CIDDS (Figure~\ref{fig:QQ-deltas-CIDDS}), TempoNet aligns closely with the diagonal across lower and mid quantiles, faithfully reproducing short-to-moderate gaps between flows. Divergence emerges in the tail, where it underproduces very long inter-arrival times. By contrast, GReaT LLM compresses spacing (inflating mid-quantiles, truncating tails), while BN\textsubscript{bins} collapses diverse timings into a narrow range due to its discretization. Overall, TempoNet better captures heavy-tailed variability, though extreme sparsity remains difficult.}

\textcolor{black}{For IoT (Figure~\ref{fig:QQ-deltas-iot}), all models struggle. \emph{BN\textsubscript{bins}} (panel a) collapses almost entirely: its quantiles remain near zero, producing a flat line instead of following the diagonal. This indicates that the baseline fails to reproduce the heavy-tailed inter-arrival distribution, instead generating unrealistically dense bursts with no long idle times.}

\textcolor{black}{\emph{GReaT LLM} (panel b) captures short inter-arrival times moderately well, with lower quantiles tracking the diagonal, but diverges sharply for longer gaps. Beyond $\sim$50 seconds, the generated values severely underrepresent the long-tail structure of IoT traffic, effectively cutting off extreme delays. In addition, we observed pathological artefacts: some samples contained inter-arrival times far beyond the valid support, with one exceeding 144,000 years. Such behavior highlights how unconstrained text-to-sequence generation can produce implausible temporal values. The net effect is unrealistic traces dominated by bursts, without the long idle periods that occur in practice when IoT devices sleep or remain inactive.}

\textcolor{black}{\emph{TempoNet} (panel c) aligns much more closely with the real distribution across the bulk of inter-arrival times (0–100 seconds), showing that it can represent both short and moderate temporal gaps. However, it truncates the extreme tail: beyond $\sim$100 seconds, the model underproduces very long idle periods. While this reduces extreme outliers, it also omits a genuine feature of IoT traffic—sporadic device activity separated by long gaps—which is essential for fully realistic simulation.}

\textcolor{black}{These plots highlight complementary failure modes across models. BN$_{\text{bins}}$ collapses temporal diversity into unrealistically dense bursts, while GReaT both truncates long gaps and occasionally produces pathological outliers (e.g., inter-arrival times on the order of 144,000 years). TempoNet avoids such extremes and reproduces realistic short-to-moderate timings more faithfully, but still under-represents the rare, very long idle periods that characterize IoT traffic. Together, these results show that while TempoNet provides the most balanced temporal fidelity, reproducing heavy-tailed sparsity across scales remains an open challenge.}

\begin{figure}[htbp!]
\includegraphics[width=\columnwidth]{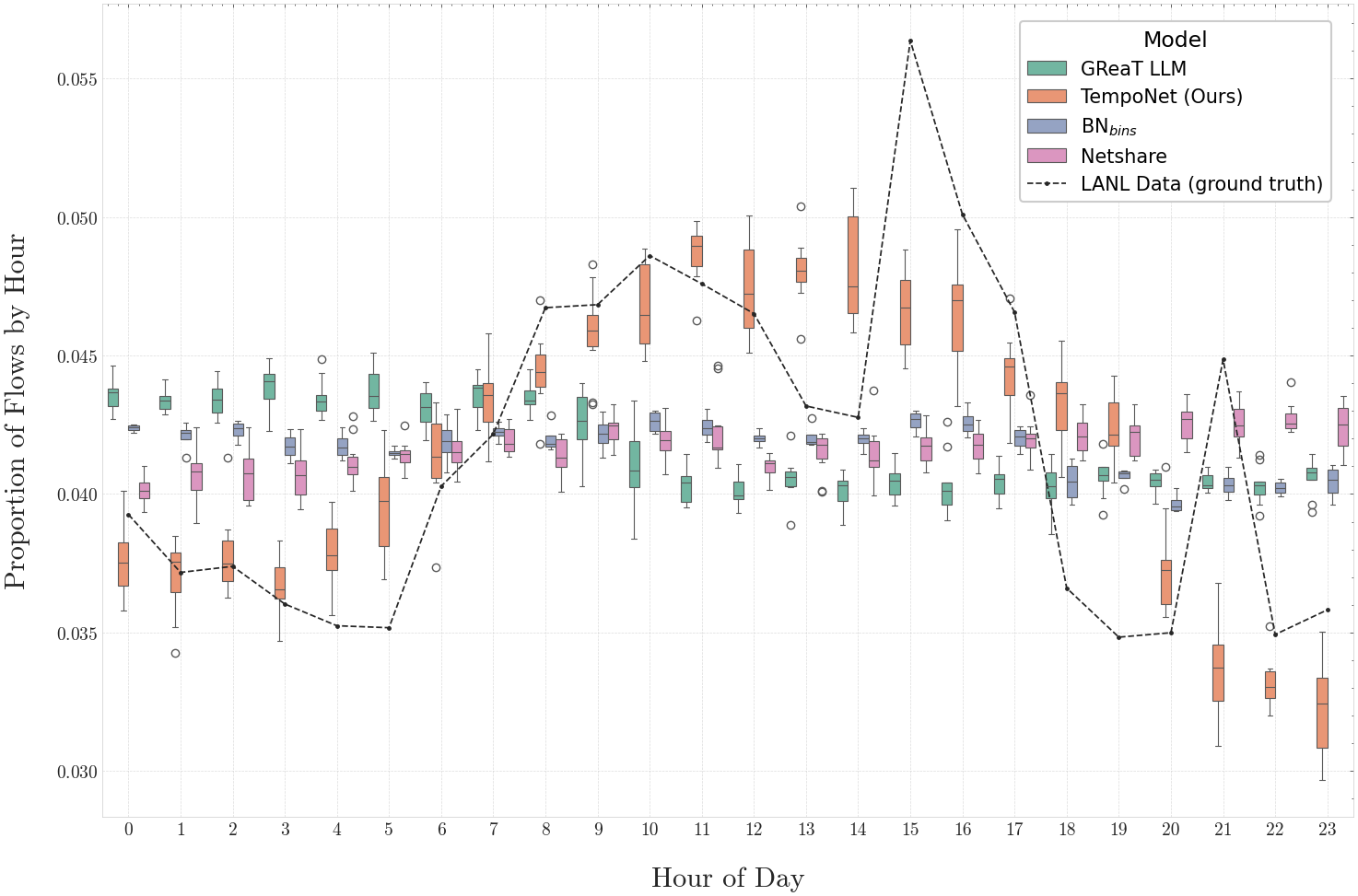}
    \caption{Daily seasonality of events in LANL dataset: Generated Vs Ground Truth.}
    \label{fig:LANL-hour-hist}
    
\includegraphics[width=\columnwidth]{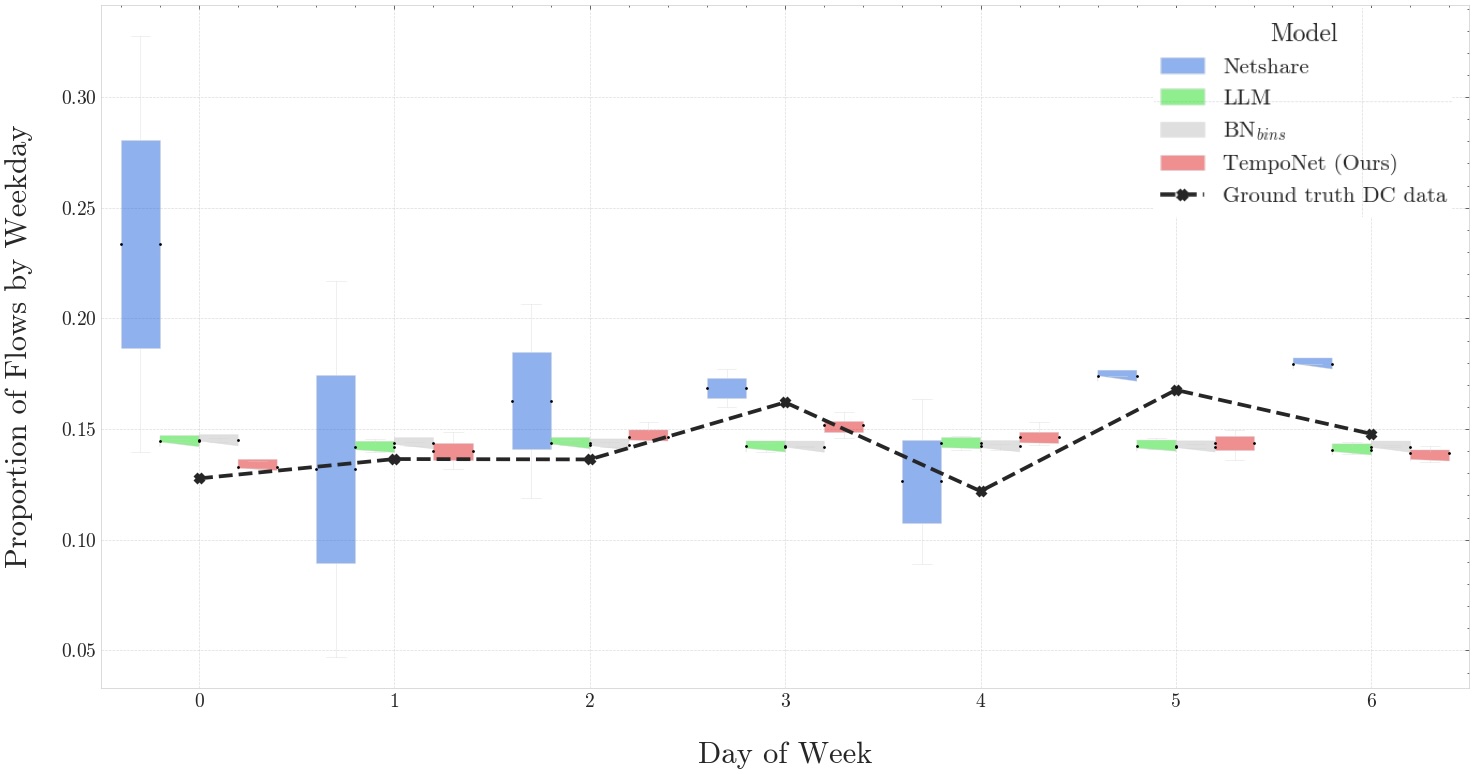}
    \caption{Weekly seasonality - DC dataset.}
    \label{fig:DC-weekday-hist}
    
\end{figure}

\begin{figure}[htbp!]
\includegraphics[width=\columnwidth]{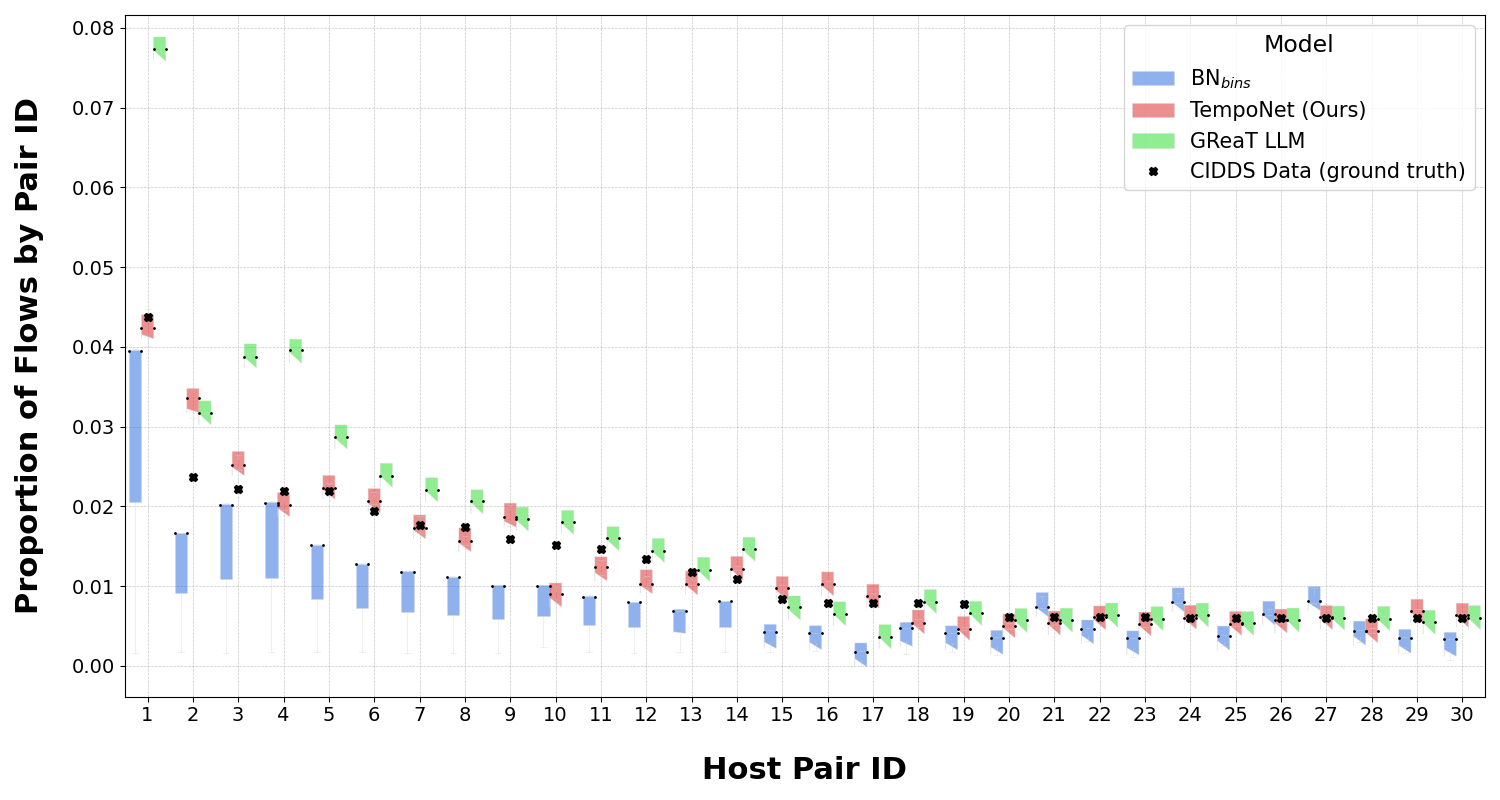}
    \caption{Host Pair Analysis - CIDDS dataset.}
    \label{fig:CIDDS-pair-hist}
\end{figure}

\begin{figure}[htbp!]
\includegraphics[width=\columnwidth]{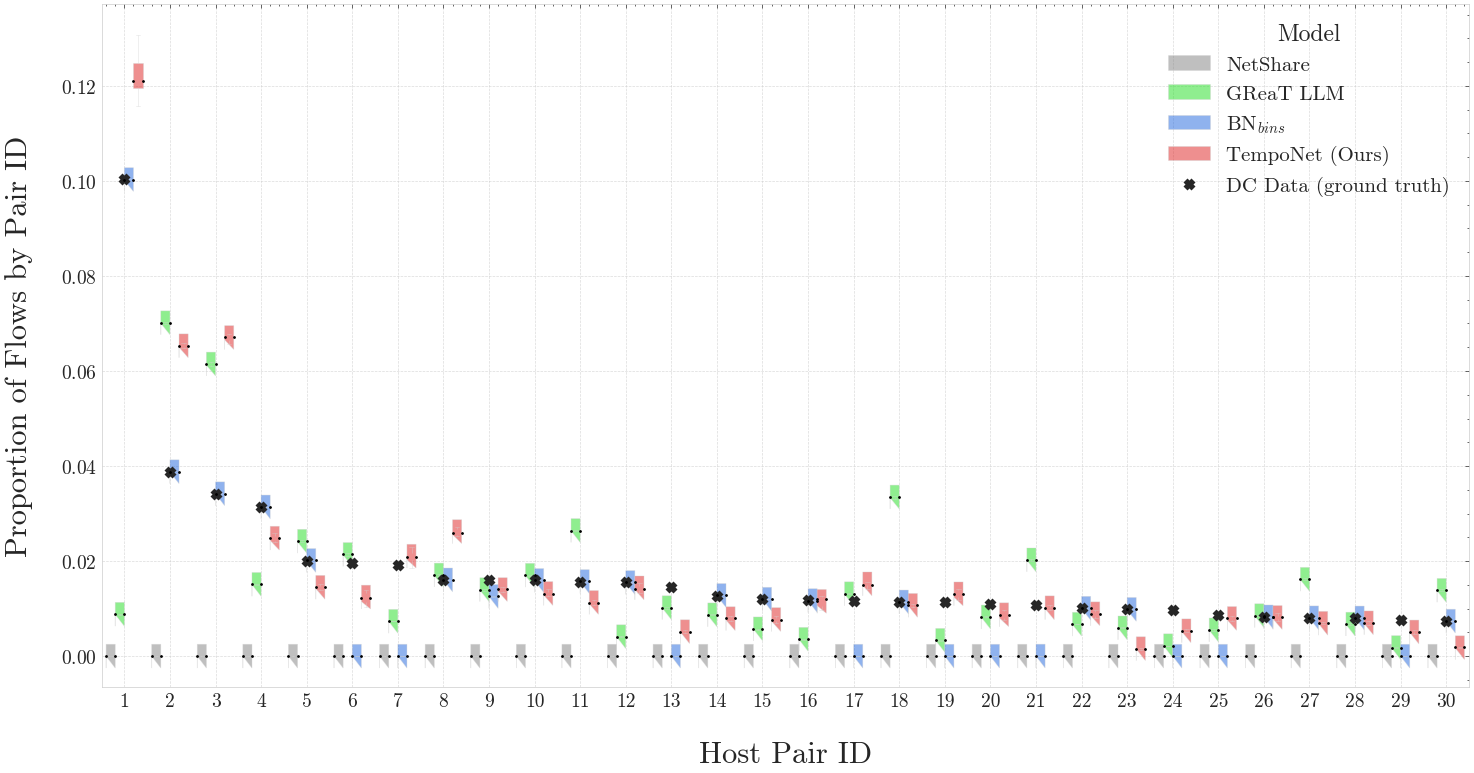}
    \caption{Host Pair Analysis - DC dataset.}
    \label{fig:DC-pair-hist}
\end{figure}

\textbf{2. Additional Visualizations.}
In addition to the temporal fidelity analysis above, we also provide the following seasonality and host-pair fidelity plots across datasets. The seasonality boxplots (Figures \ref{fig:LANL-hour-hist} and \ref{fig:DC-weekday-hist}) complement Section 6.3, confirming that TempoNet more faithfully reproduces daily and weekly cycles than either baseline. {\color{black}The host-pair fidelity plots (Figures \ref{fig:CIDDS-pair-hist} and \ref{fig:DC-pair-hist} extend Section 6.4, showing that TempoNet captures the empirical head–tail structure more faithfully than baselines. On CIDDS, BN\textsubscript{bins} systematically under-generates the most prominent source–destination pairs, while GReaT heavily over-generates them, inflating the top four pairs well beyond their true frequencies. On DC, GReaT disperses mass erratically across implausible pairs, and NetShare broadly over-disperses, flattening the natural head–tail structure. TempoNet avoids these extremes, striking a closer balance with the real distributions. These supplementary figures are consistent with the quantitative scores reported in the main text, reinforcing that the trends observed for selected datasets generalize across benchmarks.}

\subsection*{C. Compliance Check}
\label{appendix:DKC}
{\color{black}The Domain Knowledge Check (DKC) is implemented as a set of 20 lightweight plausibility rules that flag semantically or physically implausible netflow records. The rules are not intended to enforce strict protocol compliance, but rather to identify clearly unrealistic samples (e.g., HTTP traffic over UDP, flows
with payload sizes smaller than the minimum Ethernet header, or ICMP flows carrying nonzero byte counts). 

Each generated flow is evaluated against all rules, and the DKC score is computed as the proportion of rules violated. Importantly, these rules are applied \emph{post hoc} for evaluation only: the generative models have no knowledge of them during training. Even real traffic can occasionally violate some rules,
reflecting natural variability and noise in network traces.

Table~\ref{tab:dkc-rules} enumerates the full set of rules used in our evaluation. The implementation follows the public
codebase of Schoen et al.\,\cite{schoen2024tale}.}\footnote{\url{https://github.com/hallavar/SyntheticNetflowEvaluation}}

\begin{table}[ht]
\centering
\caption{{\color{black} Domain Knowledge Check (DKC) rules used to flag implausible netflow records. 
The implementation follows that Schoen et al. \cite{schoen2024tale}} 
(\url{https://github.com/hallavar/SyntheticNetflowEvaluation}).}
\label{tab:dkc-rules}
\begin{tabularx}{\columnwidth}{p{0.6cm}X}
\toprule
\textbf{\#} & \textbf{Rule Description} \\
\midrule
1  & TCP traffic on ports reserved for UDP services (53, 137, 138, 5353, 1900, 67, 3544, 8612, 3702, 123). \\
2  & UDP traffic on ports typically associated with TCP services (80, 443, 8000, 25, 993, 587, 445, 84, 8088, 8080). \\
3  & Destination port 0 with non-ICMP/IGMP traffic. \\
4  & ICMP flows with nonzero outgoing bytes. \\
5  & ICMP flows with nonzero outgoing packets. \\
6  & IGMP flows with nonzero incoming or outgoing bytes. \\
7  & NetBIOS/SSDP ports (137, 138, 1900) with nonzero incoming bytes. \\
8  & NetBIOS/SSDP ports (137, 138, 1900) with nonzero incoming packets. \\
9  & NetBIOS/SSDP ports (137, 138, 1900) not sent to broadcast addresses (.255). \\
10 & Flows targeting local/private IPs (192.168.) on external service ports (80, 443, 8000, 25, 587). \\
11 & External IPs using local-only service ports (993, 67). \\
12 & DNS port 53 flows not directed to a DNS-labeled destination. \\
13 & Multicast DNS port 5353 flows not directed to expected group addresses. \\
14 & Non-TCP flows with non-empty TCP flag fields. \\
15 & Incoming payload smaller than Ethernet header size ($\text{bytes} < 42 \times \text{packets}$). \\
16 & Outgoing payload smaller than Ethernet header size ($\text{bytes} < 42 \times \text{packets}$). \\
17 & Incoming payload larger than maximum frame size ($\text{bytes} > 65535 \times \text{packets}$). \\
18 & Outgoing payload larger than maximum frame size ($\text{bytes} > 65535 \times \text{packets}$). \\
19 & Negative or malformed durations. \\
20 & Inconsistent duration/packet relations (e.g., zero duration with multiple packets, or nonzero duration with only a single packet). \\
\bottomrule
\end{tabularx}
\end{table}
{\color{black} \textbf{DKC Violations.} TempoNet exhibits an elevated DKC violation rate on the CIDDS dataset. The most frequently violated constraint requires that flows with service ports {80, 443, 8000, 25, 587} should not target private 192.168.* addresses. While this rule encodes a sensible expectation for enterprise traffic—where such ports typically indicate external web or mail servers—it is also violated in the CIDDS training and test sets due to dataset construction artefacts. In lab-generated environments, services are often simulated within private address ranges, and in real networks, proxies or relays may legitimately run on internal hosts. TempoNet reproduces these patterns (albeit at a higher rate) rather than correcting them, as it is optimised for statistical fidelity rather than semantic validity. This highlights an important limitation of DKC evaluation: some constraints, while intuitive, are not universally valid across datasets. Future work could refine constraint sets by context or integrate such rules as soft penalties during training to reduce violations.

Beyond this primary rule, two further categories of violations emerged. First, certain violations appeared only in generated traffic: port-53 flows not mapping to DNS servers, and NetBIOS/SSDP traffic (ports 137, 138, 1900) targeting unicast rather than broadcast addresses. These cases reflect TempoNet’s imperfect capture of protocol semantics in CIDDS, as they are absent from the ground truth. Second, other violations were present in both training and generated data—for example, flows with UDP-only service ports (DNS, NetBIOS, SSDP, DHCP, NTP) incorrectly appearing with TCP as the transport protocol. Such cases point to inconsistencies in the source datasets (e.g., NetFlow artefacts or synthetic traces), which the model mirrors rather than correcting. Together, these findings illustrate how DKC evaluation can reveal both dataset–rule mismatches and genuine modelling artefacts, underscoring the need for constraint-aware generation or post-processing filters in future work.}

\subsection*{D. Expanded Related Work}

\label{appendix:related-work}

\textbf{TPPs for network and event modeling.}
Prior TPP-based models like BURSE~\cite{6782285} used Markov Modulated Poisson Processes to capture burstiness and self-similarity in workloads. Moore and Davenport~\cite{moore2016analysis} applied a multivariate Hawkes Process to trace datasets, enabling learning of topologies and change detection. NTPP~\cite{saha2019learning} integrates RMTPP~\cite{du2016recurrent} with discriminative learning-to-rank modules, modeling bursty dynamics and contention among hosts. Price-Williams and Heard~\cite{price2020nonparametric} proposed a non-parametric Wold process, outperforming Hawkes in modeling NetFlow event timing.

Beyond network traffic, TPPs have been widely applied in social and communication networks (e.g., email, Reddit, Stack Overflow)\cite{zipkin2016point, shchur2019intensity, zuo2020transformer, moore2022modelling}, and in spatio-temporal event prediction (e.g., earthquakes, disease spread, urban mobility, and neural spike trains)\cite{chenneural, zhou2022neural}. These works model event timestamps and marks, but typically support only one or two marks. In contrast, our work models seven (PCAP) or eight (NetFlow) marks per event, a significant leap in modeling complexity and application fidelity.

\textbf{GAN-based traffic models.}
NetShare~\cite{yin2022practical} builds on DoppelGANger~\cite{lin2019generating} to improve inter- and intra-epoch fidelity via fine-tuning and private training. However, NetShare does not reproduce inter-arrival time distributions, limiting its realism in time-sensitive domains. Moreover, its computational footprint is substantial—requiring 200 CPU cores and nearly 2 TB RAM across 10 Cloudlab nodes. These compute requirements pose reproducibility challenges, especially for researchers with constrained access. A notable example is the Purdue CS536 Fall 2022 class project, whose GitHub repo documents the difficulties faced in reproducing NetShare due to limited resources.

\textbf{Diffusion models.}
Recent efforts apply diffusion models to traffic synthesis, including text-to-traffic paradigms~\cite{jiang2023generative} and image-transformed flows~\cite{sivaroopan2024netdiffus}. Despite promising results, the inference speed remains a major bottleneck due to their multi-step sampling procedures. This constraint makes diffusion models impractical for real-time or high-frequency scenarios like cyber range training.

\textbf{Other models.}
Encore~\cite{huang2023datacenter} uses VAEs and GRUs to model the temporal dependency of flow sizes across host pairs. However, it only models flow sizes and does not support full header reconstruction. STAN~\cite{xu2021stan} captures intra-host dependencies using CNNs and probabilistic layers but struggles with inter-host traffic realism.

\end{document}